\documentclass[fleqn,usenatbib]{mnras}

\usepackage[T1]{fontenc}

\DeclareRobustCommand{\VAN}[3]{#2}
\let\VANthebibliography\thebibliography
\def\thebibliography{\DeclareRobustCommand{\VAN}[3]{##3}\VANthebibliography}

\usepackage{graphicx}	
\usepackage{amsmath}	
\usepackage{amssymb}	
\usepackage{bm}
\usepackage{pdflscape}

\usepackage{newtxtext,newtxmath}

\newcommand{\laplace}{\Delta}
\newcommand{\dd}{\text{d}}
\newcommand{\lb}{\lambda_{\text{dB}}}

\title[AC in SFDM halos]{Orbits and adiabatic contraction in scalar field dark matter halos: revisiting the cusp-core problem in dwarf galaxies}

\author[Pils \& Rindler-Daller]{
Kevin Pils$^{1}$\thanks{E-mail: kevin.pils@univie.ac.at}
and Tanja Rindler-Daller$^{1}$\thanks{E-mail: tanja.rindler-daller@univie.ac.at}
\\
$^{1}$Institut f\"ur Astrophysik, Universit\"atssternwarte Wien, Fakult\"at f\"ur Geowissenschaften, Geographie und Astronomie, Universit\"at Wien,\\ T\"urkenschanzstr. 17, A-1180 Vienna, Austria 
}

\date{Accepted XXX. Received YYY; in original form ZZZ}

\pubyear{2022}

\begin{document}
\label{firstpage}
\pagerange{\pageref{firstpage}--\pageref{lastpage}}
\maketitle

\begin{abstract}
Bose-Einstein-condensed dark matter, also called scalar-field dark matter (SFDM), has become a popular alternative to cold dark matter (CDM), because it predicts galactic cores, in contrast to the cusps of CDM halos ("cusp-core problem"). We continue the study of SFDM with a strong, repulsive self-interaction; the Thomas-Fermi regime of SFDM (SFDM-TF). In this model, structure formation is suppressed below a scale related to the TF radius $R_\text{TF}$, which is close to the radius of central cores in these halos. We investigate for the first time the impact of baryons onto realistic galactic SFDM-TF halo profiles by studying the process of adiabatic contraction (AC) in such halos. In doing so, we first analyse the underlying quantum Hamilton-Jacobi framework appropriate for SFDM and calculate dark matter orbits, in order to verify the validity of the assumptions usually required for AC. Then, we calculate the impact of AC onto SFDM-TF halos of mass $\sim 10^{11}~M_{\odot}$, with various baryon fractions and core radii, $R_\text{TF} \sim (0.1 - 4)$ kpc, and compare our results with observational velocity data of dwarf galaxies. We find that AC-modified SFDM-TF halos with kpc-size core radii reproduce the data well, suggesting stellar feedback may not be required. On the other hand, halos with sub-kpc core radii face the same issue than CDM, in that they are not in accordance with galaxy data in the central halo parts. 
\end{abstract}




\section{Introduction}\label{sec:Intro}


The nature of dark matter (DM) remains one of the most important open problems in astronomy and physics. The standard, collisionless cold dark matter (CDM) paradigm has been successful in explaining large-scale structure, when theoretical predictions are compared to galaxy surveys or the cosmic microwave background radiation. However, issues arise at smaller scales of (dwarf-)galaxy size, where CDM predictions fail to match with observations on these scales, referred to as the "cusp-core", "missing-satellite" and "too-big-to-fail" problems, see \cite{Bullock_small-scale-solutions} for a review. In essence, observations point to lower DM densities at the centers of host halos of dwarf galaxies ("cores"), than are predicted ("cusps"), and in fact, the cusp-core problem is of particular motivation to us in this paper. While stellar feedback in the form of central gas blowouts is considered a possible cure, the question remains open whether it is sufficient, or in accordance with properties of dwarf galaxies, such as chemical abundances.
Furthermore, the most popular CDM candidates, weakly interacting massive particles (WIMPs) as well as the QCD axion, have not yet been detected, despite decades of searches. Taken together, all these issues justify the consideration of alternative models to standard CDM, and many have been conceived.  In this paper, we consider one of these alternatives, scalar field dark matter (SFDM), which is made of (ultra-)light bosons, all condensed into their ground state, described by a single scalar field, also known as Bose-Einstein-condensed dark matter (BEC-DM). Depending on the detailed particle model, it encompasses a broad family, and the length scale, below which structure formation is suppressed, is typically much larger than that for CDM. This feature makes them attractive from the point of view of the mentioned small-scale challenges. 
If the bosons have no self-interaction (SI) and their mass $m$ is so small that their de Broglie wavelength $\lambda_\text{dB} = h/(mv)$ -- with $v$ a typical velocity close to the halo virial velocity --, is of $\sim 1$ kpc size, SFDM is known e.g. as "fuzzy dark matter (FDM)" (\cite*{Hu_SFDM}, \cite{Matos_Urena_Lopez_2000}), or "$\psi$DM" \citep{Schive_FDM}. If the particles share the same Lagrangian than the QCD axion with $m\sim 10^{-5}$ eV/c$^2$, but with much smaller mass and $\lambda_\text{dB} \sim 1$ kpc, then they are known as "ultra-light axion-like particles (ULAs)" (\cite{2010PhRvD..81l3530A}, \cite{2012PhRvD..85j3514M}). If the bosons interact via a strongly repulsive, quartic SI, when they are in the so-called Thomas-Fermi (TF) regime, SFDM has been also called "SFDM-TF" in \cite*{Dawoodbhoy_2021}, though the general terms "BEC-DM", "BEC-CDM", or "(super-)fluid DM" have been used as well; some earlier works concerning this regime include e.g. \cite{2000NewA....5..103G}, \cite{2000ApJ...534L.127P}, \cite{2007JCAP...06..025B}, \cite{rindlerdaller2012MNRAS.422..135R} and \cite{2016PDU....14...84F}.  

It is this regime of SFDM-TF which is of interest to us in this paper. In this model, a length scale due to SI arises, which is much larger than the associated $\lambda_\text{dB}$, in very contrast to the opposite case of FDM. This length scale, related to the TF radius introduced below, is the relevant scale below which structure formation is suppressed in SFDM-TF, and depending upon the particle parameters it can be of size $\gtrsim 1$ kpc.
We will be interested in studying SFDM-TF halos and their profiles. In particular, we have an eye on the cusp-core problem, i.e. the fact that observational data tends to favor cored galactic central regions, over cuspy ones which are predicted by CDM. This issue is found in dwarf galaxies of total (dynamical) masses typically around $\sim 10^{11}~M_{\odot}$. There are many theoretical approaches to study this problem, desirably with baryons included, to various levels of physical realism.
As detailed below, our paper will study and reconsider the basic assumptions which underlie the so-called method of adiabatic contraction (AC), for the first time applied to SFDM. Generally, AC is used to model the \textit{dynamic} response of DM as a result of the presence of baryons within halos by utilizing conservation laws, given certain galactic potentials. The corresponding results in terms of the final halo DM-plus-baryon profiles often match very well with more accurate (cosmological) simulations. Therefore, AC has been widely applied, given its computational simplicity and robustness.  However, in order to put our approach into perspective and to value its usefulness, we first need to review a few important findings from previous literature with respect to SFDM halos.

There have only recently been efforts to study cosmological structure and halo formation in SFDM, using DM-only simulations, some in 3D with or without realistic boundary conditions. Almost all of this body of work focuses on FDM with particle mass $m \sim (10^{-23}-10^{-22})$ eV/$c^2$, chosen such that the respective $\lambda_\text{dB} \sim (1-10)$ kpc is large enough to expose the genuine different dynamics of (S)FDM, where quantum wave phenomena, such as coherent states or wave interference, are seen at galactic scales and beyond. Owing to resolution limits, this requires relatively small simulation boxes of $\sim (1-2)$ Mpc comoving on a side; the recent work by \cite{May_Springel_2021} presents simulations up to $10$ Mpc. In the past, FDM simulations like those of \cite{Schive_FDM}, \cite*{SchwabeFDM}, \cite{Mocz_2017} and \cite*{2020arXiv200704119M} have revealed that halos, grown either in a static or expanding background, exhibit a core-envelope structure, where the coherent solitonic halo core is embedded into a highly dynamical halo envelope. Also, certain core-halo scaling relationships have been determined between core and the halo at large, although the results of simulations differ somewhat with respect to the exponents found (see also \cite{Chan_2022}), which has consequences for other observables, see e.g. \cite{Padilla_etal_2021}.  
For FDM, the solitonic core is close to the stationary ground-state solution of the Schr\"odinger-Poisson system, which are the equations of motion in this case\footnote{However, as the 1D simulations of \cite{zimmermann} show, the nature of the central object and its properties depend on how the dimensional reduction from 3D to 1D is carried out, so caution is advised. Moreover, the convergence and stability properties of isolated self-gravitating objects, with and without SI, have been studied in \cite{2004PhRvD..69l4033G}.}. On the other hand, the envelope shows CDM-like characteristics in that the outer density falloff, once properly averaged, is close to the NFW-profile, found for CDM halos (\cite*{nfw}).

While already anticipated earlier, e.g. in the papers by \cite{TRD_Shapiro_2014}, \cite{Chavanis_2019}, \cite{Padilla_etal_2021}, the actual confirmation of a core-envelope halo structure in the opposite regime of strongly-interacting, repulsive SFDM-TF has only been established very recently in \cite*{Dawoodbhoy_2021}. There, the outcome of halo formation simulations from 1D \textit{spherical} collapse starting with Jeans-unstable initial conditions in a static background has been calculated, where the combination of a 1D Lagrangian code along with a smart modelling scheme allows resolutions unattainable with 3D simulations. Core radii of order $\sim 10^{-3} - 10^{-2}$ smaller than the halo virial radius were resolved. A follow-up analysis employing more realistic cosmological boundary conditions within the same code, as well as a perturbation calculation of the linear regime of structure growth has been published subsequently in \cite*{Shapiro2021}; in physical units those core radii spanned a range of $(1 - 1000)$ pc. 

It has been found in both of these papers that SFDM-TF halo cores are close to the corresponding stationary ground-state solution of the underlying equations of motion (the Gross-Pitaevskii-Poisson system or nonlinear Schr\"odinger-Poisson system, described below); in this case the ($n=1$)-polytrope. Again, the outer envelope is CDM-like with most of its slope following an NFW-profile. (We note that the 1D results actually allow a cleaner analysis of the various sections of the calculated halo profiles, which is complicated to work out for the 3D simulations mentioned earlier and, in fact this has not nearly been worked out yet at the same degree of detail than it was for 1D). Furthermore, the work of \cite{Dawoodbhoy_2021} explains why the halo envelopes in SFDM-TF, as well as in FDM, are close to those of CDM: essentially, the novel dynamics caused by "quantum pressure" which acts on scales of $\lambda_\text{dB}$ can be coarse-grained such that it gives rise to an effective "velocity dispersion pressure" which stabilizes the halo envelope at large against gravity. The applied smoothing scale is much larger than the de Broglie scale, which is a justified procedure for the halo envelopes at large, as long as the latter are much larger than the de Broglie scale. Since this is always the case for SFDM-TF, and even true for FDM parameters of interest, it explains why both models exemplify CDM-like characteristics in their halo envelopes, at scales much larger than their respective de Broglie scale. Similar results for SFDM-TF, including a core-envelope halo structure, were found in the recent work by \cite*{Hartman_2022_simulations}, who simulated halos with $\gtrsim$ kpc-size core radii, using 3D simulations.     

Now, when it comes to the halo cores, generically found in SFDM models, a recent study by \cite*{Robles_SFDM_smallscale} has raised concerns with respect to the ability of FDM to resolve the small-scale problems, as follows. By comparing the velocity profiles of typical "too-big-to-fail halos" of masses\footnote{$M_{200}$ designates the mass of a sphere with an overdensity 200 times the ambient mean density, which results from the standard spherical collapse model in an Einstein-de-Sitter universe. The corresponding radius $R_{200}$ of that sphere is often used as a proxy for the virial radius of a galactic halo in equilibrium.} $M_{200} = 10^{9.5}-10^{10}~M_{\odot}$ in CDM and FDM to observational data of dwarf galaxies, i.e. isolated field dwarfs and members of the Local Group that are not satellites, they find that FDM halos can fit the data well, while CDM overpredicts central densities, as expected. However, when doing the same comparison with typical "cusp-core halos" which are more massive, $M_{200} = 10^{10}-10^{11.15}~M_{\odot}$, and using the same FDM parameters than before, then FDM clearly fails in explaining the data. 
Subsequently, \cite{Dawoodbhoy_2021} also performed such a comparison, but including SFDM-TF as a further model. Indeed, while they confirm that CDM fails, and FDM fails with respect to the cusp-core halos, SFDM-TF halos can explain the data of \textit{both} halo mass ranges, \textit{if} the TF radius of the central core of those halos is large enough; e.g. $4$ kpc size. The basic reason stems from the different ways of how the central densities scale in these models; for SFDM-TF the core densities rise mildly with core mass, in contrast to FDM. As a result, SFDM-TF fares better than FDM in both small-scale issues.

These are interesting findings with implications for the reliability of SFDM as an alternative to CDM. However, all these papers mentioned so far were based on DM-only studies, and the dynamic impact of the baryonic component of halos was neglected.

There is only one paper where FDM simulations \textit{with the addition} of baryons have been presented, namely \cite{Mocz2020}, and no DM-plus-baryon simulations for SFDM-TF have been performed, so for comparison's sake, we will put our work here into context to some of the results from \cite{Mocz2020} in the forthcoming.
In that paper, similar to previous cosmological FDM simulations, a boson mass of $\sim 10^{-22}$ eV/$c^2$ is chosen and the comoving simulation box is small from a cosmological perspective, $\sim 2$ Mpc on a side. However, this new simulation includes a host of implemented baryon physics, like star formation, supernova feedback, cooling and (instantaneous) reionization at a redshift of $z \sim 6$. Over time the resolution of the de Broglie scale becomes increasingly computationally prohibitive, therefore the cosmological simulation is stopped at $z = 5.5$. Again, wave interference is found, not only within halos, but also stretching into the cosmic filaments connecting these halos, as well as the coherent solitonic cores at halo centers, with radius $\sim 1$ kpc at $z \sim 6$. In fact, \cite{Mocz2020} characterize the three halos, each with $M_{200} \sim 10^{10} ~M_{\odot}$, that they find at the end of their simulation run, which stops too early to make any direct connection to present-day galaxies. However, we will use the most massive halo of their simulation as one comparison to our AC-modified SFDM-plus-baryon halo profiles presented below.

Given the complications of performing and interpreting fully-fledged cosmological structure formation simulations of SFDM universes, and the demands on resolution that will not be overcome any time soon, it is important to approach the problem of SFDM halo structure and formation from multiple angles. In particular, we deem it necessary to improve our analytic understanding and to come up with theoretically motivated models which can be handled easily enough that a comparison to galaxy data is practicable. While this comparison is important in any case, we have especially the cusp-core problem in mind as a motivation to study the question of whether SFDM can be a cure, or not. 

A first thing to consider, when adding baryons to DM halos, is the phenomenon of gravitational contraction, i.e. the dynamic response of DM due to the gravity by infalling baryons. If the gravitational potential changes slowly, compared to typical orbital frequencies, it is possible to use so-called "adiabatic invariants", hence the term adiabatic contraction ("AC"), which are based on integrals of motion. Among the first to use this approach, even before the CDM paradigm had been established, were \cite*{Eggen1962}, \citet{Zeldovich1980} and \citet{BarnesWhite}. However, the most influential publications on this topic were \citet{blumenthal1986} and \citet{RydenGunn}, whose studies were carried out concurrently. They used AC as a means to explain flat galactic rotation curves, also applied to the Milky Way. Ever since then, AC has been used in order to study galactic halo profiles, although improvements of the original approach have been devised, in particular by the work of \citet{Gnedin_2004}.

In this paper, we will study SFDM-plus-baryon halo profiles and how they fare with respect to typical "cusp-core halos" within the framework of AC. Before that, we need to revisit the fundamental assumptions upon which AC is based, in order to determine whether and how it can be applied to SFDM.
Other than the dynamical impact, we will not include further baryon feedback to our models. While pure AC may seem insufficient in order to compare to observations, we reiterate that it is a first important step. In fact, even AC-modified density profiles have not regularly been applied, when it comes to the comparison to galaxy data, a shortcoming that has been recently pointed out by \cite{Cautun2020} in the context of the Milky Way. For one thing, the incorporation of AC yields upper-bound estimates of the densities and velocities within SFDM halos, once baryons are included. Thereby, our work can help to assess how much stellar feedback is actually allowed or required, in order to bring SFDM models in alignment with galaxy data.  

Although we neglect a good deal of complicated astrophysics, our task at hand is not simple, either. More precisely, our work was motivated by three fundamental questions: i) Given that SFDM exemplifies quantum behavior, in which sense can we talk about "dark matter orbits"? Or rephrased, can we understand orbits within the framework of the quantum Hamilton-Jacobi equation that extends the classical approach to the quantum realm, where we encounter "quantum trajectories"? ii) If the notion of orbits is extended, can we apply the AC methodology used from collisionless CDM to SFDM? iii) If AC can be properly modified and used, what is the impact of baryons within SFDM halos that we will find thereafter, compared to CDM? Or rephrased, will SFDM halos \textit{with} baryons added be able to resolve the cusp-core problem? 

We were able to answer i) and ii) in a positive manner for SFDM-TF, the results of which will be presented in this paper. Also, we present our results with respect to question iii), where we identify the required properties of SFDM-TF in order to make a cure to the cusp-core problem.  Work is in progress to study these questions also for the opposite regime of FDM, which will be presented elsewhere.

This paper is organized as follows: Section \ref{sec2} includes a short recap of the fundamental equations of motion underlying SFDM. In Section \ref{sec3}, we present the TF regime of SFDM in more detail, including the core-envelope halo structure that we will employ. In Section \ref{sec4}, we derive the quantum Hamilton-Jacobi equation, including the calculation of some example orbits within the TF potential that describes the halo core. Section \ref{sec:adiabatic-contraction} contains our application of the AC methodology to SFDM halos, in which we will also compare our results to observational velocity data of dwarf galaxies. Finally, Section \ref{sec:conclude} summarizes our main results and conclusions.

\section{Fundamental equations of self-gravitating SFDM/BEC-DM halos}
\label{sec2}

We assume that SFDM consists of a single species of bosons with particle mass $m$, whose dynamics can be described by a complex function $\psi(\mathbfit{r}, t)$, which is basically the "wave function of the condensate" of the BEC, formed by these bosons.
As in standard CDM, galactic SFDM halos are nonlinear overdensities, compared to the background of a $\Lambda$SFDM universe which is assumed to be homogeneous and isotropic as in $\Lambda$CDM.

We are interested in the dynamical description of (individual) halos, whose evolution and structure can be described by the  
Gross-Pitaevskii (GP) equation (see \cite{1961NCim...20..454G,pitaevskii1961vortex}, and applied to gravity in \cite{1968PhRv..172.1331K,1969PhRv..187.1767R}),
\begin{align} \label{eq:gross-pitaevskii}
    i \hbar \frac{\partial \psi(\mathbfit{r}, t)}{\partial t} = -\frac{\hbar^2}{2m} \laplace \psi(\mathbfit{r}, t) + \left(m\Phi(\mathbfit{r},t) + g|\psi(\mathbfit{r}, t)|^2 \right) \psi(\mathbfit{r}, t),
\end{align}
which is coupled to the Poisson equation
\begin{align} \label{eq:poisson}
    \laplace \Phi(\mathbfit{r},t) = 4\pi G m |\psi(\mathbfit{r}, t)|^2.
\end{align}
The full system is called Gross-Pitaevskii-Poisson (GPP) equations.
The Born assignment is used, such that $|\psi(\mathbfit{r}, t)|^2 = n(\mathbfit{r}, t)$ describes the number probability density of the bosons.
The $2$-boson contact self-interaction (SI) is modelled as the third term on the right-hand side of (\ref{eq:gross-pitaevskii}), where $g$ is a constant coupling strength that determines whether the bosons interact attractively ($g < 0$), or repulsively ($g > 0$). Together with $m$, it is a free parameter of the SFDM model. $\Phi$ is the gravitational potential of the self-gravitating halo and enters in the 
GP equation like an external trapping potential would do for BEC laboratory systems. In addition to the nonlinear SI-term in the GP equation, the full GPP equations are nonlinear in any case due to their coupling, even if we set $g=0$. This case corresponds to FDM models, discussed in the Introduction.  

The assumption that all bosons within a given halo of volume $V$ can be described by $\psi$ then naturally leads to the normalization condition,
\begin{align*}
    \int_V |\psi|^2 = N.
\end{align*}

The literature has made extensive use of an equivalent representation of GPP, \eqref{eq:gross-pitaevskii} and \eqref{eq:poisson}, by transforming the GP equation into quantum hydrodynamic form, pioneered in particularly by \cite{Bohm52-1,Bohm52-2} and \cite{Taka54}.
Using the polar decomposition or Madelung transformation \citep{madelung}, the wave function is decomposed into its phase and amplitude functions, 
\begin{align} \label{eq:madelung}
    \psi(\mathbfit{r}, t) = |\psi(\mathbfit{r}, t)|e^{iS(\mathbfit{r}, t)/\hbar} = \sqrt{\frac{\rho(\mathbfit{r}, t)}{m}} e^{iS(\mathbfit{r}, t)/\hbar}.
\end{align}
Here we identify $\rho(\mathbfit{r}, t)$ as the SFDM halo mass density and $S(\mathbfit{r}, t)$ as the action function. It is related to the associated bulk velocity of the halo in this representation\footnote{In much of the literature, $\hbar$ is absorbed in the definition of $S$ in (\ref{eq:madelung}), resulting in the appearance of $\hbar$ in the numerator of equation (\ref{eq:bulk-velocity}). However, our choice of definition here is more appropriate for the quantum Hamilton-Jacobi framework that we apply in this paper.} as follows,
\begin{align} \label{eq:bulk-velocity}
    \mathbfit{v} = \frac{1}{m} \nabla S.
\end{align}
Consequently, substitution of equation \eqref{eq:madelung} into the GP equation \eqref{eq:gross-pitaevskii} yields, after separating the real and imaginary parts, 
\begin{align} \label{eq:gp-real}
    -|\psi|^2 \frac{\partial S}{\partial t} + \frac{\hbar^2}{2m} |\psi| \laplace |\psi| - \frac{1}{2m} |\psi|^2 (\nabla S)^2 - \left(m\Phi + g|\psi|^2 \right) |\psi|^2 = 0
\end{align}
and
\begin{align} \label{eq:gp-imaginary}
    \frac{\partial |\psi|^2}{\partial t} + \nabla \cdot \left( \frac{1}{m} |\psi|^2 \nabla S \right) = 0.
\end{align}
Making use of the bulk velocity, these equations are then further transformed into a set of hydrodynamic equations, consisting of a continuity equation and an Euler-like momentum equation,
\begin{align}
    &\frac{\partial \rho}{\partial t} + \nabla \cdot (\rho \mathbfit{v}) = 0, \label{eq:continuity} \\
    &\frac{\partial \mathbfit{v}}{\partial t} + \left( \mathbfit{v} \cdot \nabla \right) \mathbfit{v} = - \nabla Q - \nabla \Phi - \frac{1}{\rho} \nabla P_\text{si}, \label{eq:euler-eq} \\
    &\laplace \Phi = 4 \pi G \rho, \label{eq:poisson2}
\end{align}
which are supplemented by the Poisson equation (\ref{eq:poisson2}).
The involved quantities are well-known in the field, namely the so-called quantum or Bohm potential,
\begin{align} \label{eq:quantum-potential}
    Q = -\frac{\hbar^2}{2m^2} \frac{\laplace \sqrt{\rho}}{\sqrt{\rho}},
\end{align}
and a pressure of polytropic form with index\footnote{This notation shall not be confused with the number density $n(\mathbfit{r},t)$; the distinction should be clear in context.} $n=1$,
\begin{align} \label{eq:polytropic-pressure}
    P_\text{si} = K_\rho \rho^{1 + 1/n} = \frac{g}{2m^2} \rho^2,
\end{align}
which originates from the SI between bosons (see e.g. \cite{rindlerdaller2012MNRAS.422..135R} for more details on the respective limit cases of the fundamental equations).

\section{The Thomas-Fermi regime of strongly repulsive self-interaction: SFDM-TF}
\label{sec3}

In principle, one could go ahead and solve the system of equations (\ref{eq:continuity})-(\ref{eq:poisson2}) for any given initial and boundary conditions of a halo in question, and add the baryonic component and its equations on top of it. This is a very complicated task, even for an isolated halo in a static, non-expanding background universe. Also, a goal of ours in this paper concerns a better understanding of the orbital nature and structure that we can expect in SFDM halos, as well as the impact of baryons onto halo profiles. For these reasons, we will employ a "semi-analytic" modelling approach, where we include the SI-pressure in (\ref{eq:polytropic-pressure}) in an exact manner, but approximate the physics of the $Q$-term in (\ref{eq:quantum-potential}) in a coarse-grained way.
The fundamental formalism of this approach has been derived and laid out in detail in \citet{Dawoodbhoy_2021}, whose main results we discussed already in the Introduction.
Importantly, this approach works for SFDM models with strongly repulsive SI, thus "SFDM-TF", i.e. for cases where the SI-related length scale is much larger than the de Broglie length $\lb$, making it effectively the only relevant scale in the system.  In fact, the aforementioned coarse-graining uses a smoothing scale that is much larger than $\lb$; as a result we cannot resolve scales around or below $\lb$. However, it turns out that this is no shortcoming at all, if we are interested in modelling and understanding gross quantities like density profiles and their evolution, for halos in these SFDM-TF models.  

For the purpose of the work in this paper, we will only use here one of the "semi-analytic" type of models of \citet{Dawoodbhoy_2021} which was mostly applied for the sake of comparison to the simulated halos. However, it has been shown there that this model is a good approximation for the accurate SFDM-TF halo profiles, and we will summarize it below. Interestingly, the 3D simulation results of \cite{Hartman_2022_simulations} also seem to be in good accordance with these semi-analytic models, strengthening their viability as good approximations for halo structure in SFDM-TF. 

Basically, in this model the \textit{halo cores} are approximated by the static ground-state solution of GPP, or equivalently (\ref{eq:continuity})-(\ref{eq:poisson2}), for which $\hbar$ is set to zero, which means there is no $Q$-term. This limit case has been studied previously at quite some detail, see e.g. \cite{rindlerdaller2012MNRAS.422..135R}. It is usually a good model only for the cores of SFDM-TF halos; we call them Thomas-Fermi (TF) cores, because they constitute the minimum-size objects that can exist in virial equilibrium. As such, they might represent the smallest halos in SFDM-TF models, which should reasonably host the smallest galaxies, in turn.

On the other hand, for halos that host large galaxies and that are larger than these minimum-size objects, they have to have extended halo envelopes, as mentioned in the Introduction. These envelopes stem from the complicated dynamics mediated by the $Q$-term. As shown in \citet{Dawoodbhoy_2021}, the overall effect of this large-scale dynamics can be modelled like a "classic" velocity dispersion $\sigma^2$, which gives rise to a macroscopic "velocity dispersion pressure", $P_{\sigma}$, very similar\footnote{Another analogue is the velocity dispersion of CDM particles, using momentum moments of the collisionless Boltzmann equation (also Vlasov-Poisson equations).} to the stellar velocity dispersion in elliptical galaxies, say, which enters the respective Jeans equations.   
This halo envelope is then "stitched together" with the central TF halo core, in order to provide a complete model for a given SFDM-TF halo. Baryons are then included into this model in Section \ref{sec:adiabatic-contraction}.  Before we get there, we describe the dark matter component first, though we recommend that the reader consult \citet{Dawoodbhoy_2021} for details.

\subsection{TF core}
\label{sec31}

Let us now set $\hbar = 0$ and consider the static case; 
the Euler-like equation \eqref{eq:euler-eq} reduces to
\begin{align} \label{eq:TF-regime-diffeq}
    -\frac{1}{\rho} \nabla P_\text{si} = \nabla \Phi
\end{align}
and only SI counteracts gravity. In spherical symmetry, it can be shown that this equation leads to the well-known Lane-Emden equation for an $(n=1)$-polytrope, whose solution for the density profile is given by
\begin{align} \label{eq:polytrope-density}
    \rho(r) = \rho_0 \frac{\sin (\pi r / R_\text{TF})}{\pi r / R_\text{TF}},
\end{align}
with central density $\rho_0$, and $R_\text{TF}$ is the first zero of the density profile, which serves as the radius of the polytrope, the so-called TF radius. It is given by
\begin{align} \label{eq:tfradius}
    R_\text{TF} = \pi \sqrt{\frac{g}{4 \pi G m^2}},
\end{align}
i.e. it is fully determined by a given choice of the SFDM parameter \textit{combination}, $g/m^2$, and it does not depend on the total mass of the polytrope. That is, in the context of SFDM-TF halos, their core size does not depend on the core mass. Hence, once we specify a SFDM model by fixing $g/m^2$, we get a fixed value for $R_\text{TF}$, independent of core or total mass. Conversely, if galaxy observations demand upper bounds on the core radius $R_\text{TF}$, this implies upper bounds on $g/m^2$ and thus constraints on SFDM-TF models. 

Using the Poisson equation (\ref{eq:poisson2}) in spherical symmetry, the gravitational potential can be calculated from the density in eq. \eqref{eq:polytrope-density}, resulting in
\begin{align} \label{eq:polytrope-gravpot}
    \Phi(r) = - \frac{4 G \rho_0 R_\text{TF}^3}{\pi ^2} \frac{\sin(\pi r / R_\text{TF})}{r}.
\end{align}
These calculations are not new, but to be self-contained, we provide in Appendix \ref{appendix:thomas-fermi} derivations of the density profile \eqref{eq:polytrope-density}, the gravitational potential \eqref{eq:polytrope-gravpot}, as well as analytical expressions for the mass profile, $M(r)$, and the circular velocity profile, $v(r)$, for the TF core.

\subsection{Core-envelope halo structure} \label{subsec:core-envelope}

A pure, single TF core is held up in gravitational equilibrium due to the SI-pressure, $P_{\text{si}}$, in equation (\ref{eq:TF-regime-diffeq}). On the other hand, as just described above, there is an envelope which surrounds this core, and which is stabilized due to velocity dispersion $\sigma^2$, giving rise to a velocity dispersion pressure, $P_{\sigma}$, which is a coarse-grained description of the more complicated dynamics mediated by the $Q$-term. However, this still leaves the question open which relationship between $\sigma^2$ and $P_{\sigma}$ one should adopt. We assume that $\sigma^2$ is a constant, independent of radius, i.e. we set
\begin{align} \label{eq:velocity-dispersion}
    \sigma^2 = \frac{P_\sigma}{\rho} = \text{constant.}
\end{align}
Thus, the pressure of the envelope is that of an isothermal sphere (again, stressing that this is an effective description), which is useful for computational simplicity, and reflects the fact that CDM halos have been also described with isothermal profiles in the past; this way we follow the same reasoning than \citet{Dawoodbhoy_2021} in our choice of $\sigma^2$. 
Since an isothermal sphere can be described as a polytrope with index $n=\infty$, the combined model of ($n=1$)-polytropic TF core plus isothermal envelope can be considered a "double-polytrope".  One can readily write down the differential equation which describes hydrostatic equilibrium for this halo model in spherical symmetry, 
\begin{equation}\label{eq:ce}
    -\frac{1}{\rho} \frac{\dd }{\dd r} \left( \sigma^2 \rho + \frac{g}{2m^2} \rho^2 \right) = \frac{\dd \Phi}{\dd r},
\end{equation}
see \citet{Chavanis_2019} and \citet{Dawoodbhoy_2021} for details.
It also includes the two limit cases that it embodies, namely if $\sigma=0$ we have only the self-gravitating, spherically-symmetric TF core through (\ref{eq:TF-regime-diffeq}), or else if $g=0=P_{\text{si}}$, i.e. if there is no SI, we have a self-gravitating, purely isothermal, spherically-symmetric system with pressure $P_{\sigma} = \rho \sigma^2$, without a TF core.
We stress again that this halo model produces density profiles that are fairly good approximations of the exact profiles found by \citet{Dawoodbhoy_2021}, and their computational simplicity is beneficial to our studies here.


Before we proceed, we take the divergence of both sides of (\ref{eq:ce}), multiply by $r^2$ and use the Poisson equation, leading to
\begin{align} \label{eq:core_envelope_1}
    -\frac{\dd}{\dd r} \left( \sigma^2 r^2 \frac{\dd \ln \rho}{\dd r} + \frac{g}{m^2} r^2 \frac{\dd \rho}{\dd r} \right) = 4 \pi G r^2 \rho.
\end{align}
There are several ways of non-dimensionalizing this equation. We follow the approach by \citet{Chavanis_2019}, which is also applied in \citet{Dawoodbhoy_2021}, which is based on the usual ansatz to come up with the Lane-Emden equation. Thus, we substitute
\begin{align*}
    \rho = \rho_0 e^{-\Xi} \quad \quad \text{and} \quad \quad \xi = \frac{r}{r_0},
\end{align*}
where
\begin{align}
    r_0 = \left( \frac{\sigma^2}{4 \pi G \rho_0} \right)^{1/2}
\end{align}
is the characteristic radius of the isothermal sphere. Furthermore, a parameter is defined which characterizes the strength of the SI-polytropic pressure over the velocity dispersion pressure modelled by $\sigma$, i.e.
\begin{align} \label{eq:chi_parameter}
    \chi = \frac{\pi^2 g \rho_0}{m^2 \sigma^2} = \left( \frac{R_\text{TF}}{\pi r_0} \right)^2.
\end{align}
Utilizing these expressions, the non-dimensionalized version of equation \eqref{eq:core_envelope_1} reads as
\begin{align} \label{eq:core_envelope_2}
    \frac{1}{\xi^2} \frac{\dd }{\dd \xi} \left( \xi^2 \frac{\dd \Xi}{\dd \xi} + \chi \xi^2 e^{-\Xi} \frac{\dd \Xi}{\dd \xi} \right) = e^{-\Xi}.
\end{align}
Now, this differential equation unifies two limit cases: if $\chi = 0$ (eqivalent to $R_\text{TF} = 0$, i.e. no SI) we have the Emden-Chandrasekhar equation for an isothermal sphere. On the other hand, for large $\chi$ and neglect of the first term in (\ref{eq:core_envelope_2}), we get the Lane-Emden equation for an ($n=1$)-polytrope.
We accompany equation (\ref{eq:core_envelope_2}) with the boundary conditions
\begin{align} \label{eq:BC}
    \Xi(0) = 0 = \Xi'(0),
\end{align}
in order to ensure that our halo density profiles always have a finite central density.

\begin{figure}
    \centering
    \includegraphics[width=\columnwidth]{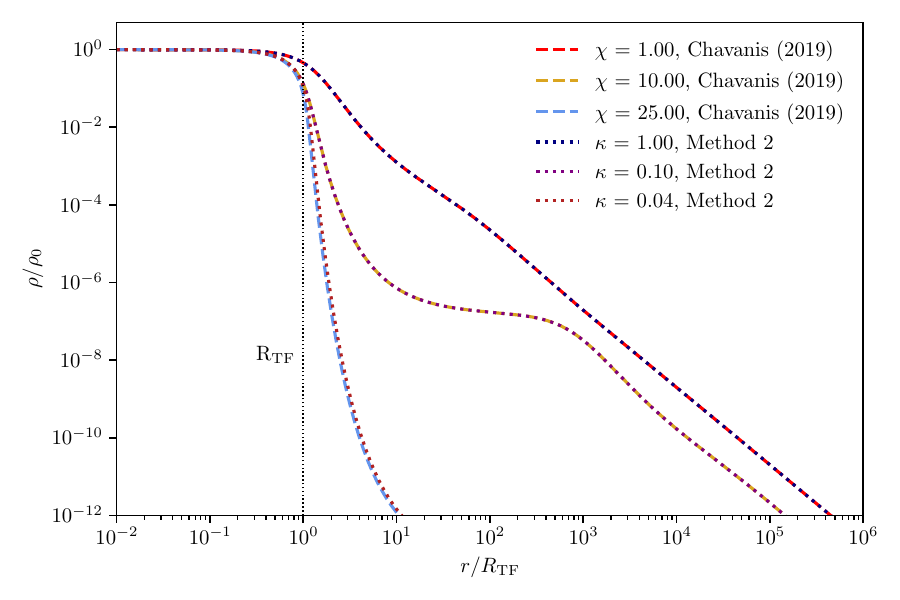}
    \caption{\textbf{Density profiles for a core-envelope SFDM-TF halo according to the double-polytrope model} as obtained from equations \eqref{eq:core_envelope_2} (dashed curves) and \eqref{eq:core_envelope_3} (dotted curves), respectively. The comparison shows identical results for sufficiently small $\chi$ (equivalent to sufficiently large $\kappa$), but some deviations occur for larger $\chi$ (smaller $\kappa$). The parameter $\chi (=1/\kappa)$ describes the dominance of the TF core over the envelope.} 
    \label{fig:core-envelope-comparison}
\end{figure}

In Figure \ref{fig:core-envelope-comparison}, we plot the solutions of equation (\ref{eq:core_envelope_2})-(\ref{eq:BC}) for different choice of $\chi$ (dashed curves). The larger $\chi$, the more dominant the TF core over the envelope becomes, which increasingly "shrinks", as a result. A pronounced transition between core and envelope occurs for intermediate values of $\chi$. Obviously, this feature is subject to observational constraints. For halos with a mild enough transition between core and envelope, a value of $\chi$ close to $1$ is required; see \citet{Dawoodbhoy_2021}. 

For comparison, we present a second, equivalent version of non-dimensionalizing equation \eqref{eq:core_envelope_1} in Appendix \ref{appendix:core_envelope_method_2}. Figure \ref{fig:core-envelope-comparison} also shows plots of the respective solutions of that second method (dotted curves). The corresponding parameter in this case is $\kappa = \sigma^2/v_c^2 = 1/\chi$, see Appendix \ref{appendix:core_envelope_method_2}. While the two methods agree in the solution over most of the radial range, we only use the first one which led to (\ref{eq:core_envelope_2}) in our forthcoming analysis, since it yields numerically more stable results.

The SFDM-TF halo profiles that are solutions to (\ref{eq:core_envelope_2})-(\ref{eq:BC}) will enter our AC routine below, as our profiles of choice when we have to specify the initial dark matter distribution. But first, we have to study the notion of orbits that we encounter in the TF cores of our halos, given the nature of the equations of motion in Section \ref{sec2}. To this end, it will be useful to employ the Hamilton-Jacobi framework, described and applied in the next section.

\section{Quantum-Hamilton-Jacobi equation and orbits in the TF core}
\label{sec4}

An equivalent and useful formulation of SFDM, apart from the quantum hydrodynamical equations presented in Section \ref{sec2}, is given by the quantum Hamilton-Jacobi equation in (\ref{eq:hamiltonian}) below, which arises naturally from the GP equation (\ref{eq:gross-pitaevskii}) upon the Madelung transformation (\ref{eq:gp-real})-(\ref{eq:gp-imaginary}), see \citet{Wyatt} for a textbook presentation. First, a Hamiltonian $\mathcal{H}$ can be readily derived
\begin{align} \label{eq:hamiltonian}
    \mathcal{H} = \frac{1}{2m}(\nabla S(\mathbfit{r},t))^2 + mQ(\mathbfit{r},t) + m\Phi(\mathbfit{r},t) + \frac{g\rho(\mathbfit{r},t)}{m} = -\frac{\partial S}{\partial t}.
\end{align}
The left-hand equality can be made upon realizing that the term $(\nabla S)^2/2m = m\mathbfit{v}^2/2 = \mathbfit{p}^2/2m$ is the kinetic energy of our system, while the other terms constitute the quantum, gravitational and SI potential energy, respectively.  Looking now at the right-hand equality, we recognize (\ref{eq:hamiltonian}) as a quantum Hamilton-Jacobi equation (QHJE), in an Eulerian frame of reference. As usual, the function in question, $S$, can be regarded as an action,
and is also referred to as (quantum) Hamilton's principle function.
Now, the appearance of the quantum potential $Q$, which gives rise to a "quantum force" in equation (\ref{eq:euler-eq}), affects the trajectories of our mechanical system. $Q$ introduces non-locality and as a result trajectories become correlated; they are called quantum trajectories. 
The QHJE can be used as an alternative route to solve the dynamics of the system in question, enabling the search for canonical transformations to find integrals of motion, upon which the analysis can be possibly simplified. The quantum potential $Q$ is generally time-dependent, because the density $\rho$ (and amplitude of the wave function) are time-dependent when evaluated along a quantum trajectory. Consequently, the total energy evaluated along such a trajectory is not constant. However, we will make the assumption that our galactic density-gravitational-potential pairs are time-independent, i.e. we assume that $\Phi = \Phi(\mathbfit{r})$ and $\rho = \rho(\mathbfit{r})$. This way, the quantum and the SI potentials are time-independent, as well. We also consider spherical coordinates $\mathbfit{q} = (r, \vartheta, \varphi)$ and $\mathbfit{p} = (p_r, p_{\vartheta}, p_{\varphi})$. In fact, as seen from (\ref{eq:hamiltonian}) along with the kinetic energy, $\mathbfit{q}$ and $\mathbfit{p}$ can be already identified as canonically conjugated variables as follows,
\begin{align}
    p_r &= \partial S/\partial r, \label{eq:momentum-r} \\
    p_\vartheta &= \partial S/\partial \vartheta, \label{eq:momentum-theta}\\
    p_\varphi &= \partial S/\partial \varphi. \label{eq:momentum-phi}
\end{align}
With this connection, and looking at equation \eqref{eq:hamiltonian},
we can immediately identify a new Hamiltonian that is zero, $\mathcal{K} = \mathcal{H} + \partial S/\partial t = 0$, i.e. it is constant and independent of time. We merely rewrite \eqref{eq:hamiltonian} in the form
\begin{align} \label{eq:qhje}
    \mathcal{K} \left(\mathbfit{r}, \frac{\partial S}{\partial \mathbfit{r}}, t \right) &= \frac{1}{2m} \left( \frac{\partial S}{\partial r} \right)^2 + mQ(r,\vartheta,\varphi) + m\Phi(r,\vartheta,\varphi) + \frac{g}{m}\rho(r,\vartheta,\varphi) \nonumber \\ 
    &+ \frac{1}{2mr^2} \left( \frac{\partial S}{\partial \vartheta} \right)^2 + \frac{1}{2mr^2 \sin^2 \vartheta} \left( \frac{\partial S}{\partial \varphi} \right)^2 + \frac{\partial S}{\partial t} = 0,
\end{align}
where $S$ depends on all coordinates and time, resulting in the desired QHJE. It is still hard to solve this partial differential equation, unless a separation of variables is feasible. Whether this is the case in general depends on the chosen coordinate system, as well as the physical problem itself. Even though there is no general way of finding out whether separability can be fully applied, the so-called St\"ackel conditions (developed by \cite{staeckel}; see also \citet{bookGoldstein}, \citet{boccaletti}) provide a useful tool for orthogonal coordinate systems. The conditions, as well as the calculations necessary to check them, are presented in Appendix \ref{appendix:staeckel-conditions}.
In doing so, we make another assumption, namely that the density-gravitational potential pairs are spherically-symmetric, i.e. $\Phi = \Phi(r), \rho=\rho(r)$, while $\mathbfit{v}$ - and hence $\mathbfit{p}$ and $S$ - can depend also on angular variables. Consequently, $S$ will have the form of (\ref{eq:principle-function}),
$S(\mathbfit{q},\mathbfit{p},t) = W(\mathbfit{q},\mathbfit{p}) - Et$, with constant momenta $p_i$ and $\dd W/\dd t = p_i \dot{q_i}$, or $W=\int p_i \dd q_i$, where $i=r,\vartheta,\varphi$.
Then, the following integrals of motion are identified (denoted $\alpha_i$ in Appendix \ref{appendix:staeckel-conditions}): 
\begin{enumerate} 
    \item energy, $\mathcal{H} = \alpha_t = E$
    \item angular momentum component in $z$-direction, $p_\varphi = \alpha_\varphi = L_z$
    \item magnitude of the angular momentum vector, $\alpha_\vartheta = |\mathbfit{L}| = L$, or in terms of the momentum, $p_\vartheta = (L^2 - p_\varphi^2/\sin^2\vartheta)^{1/2}$
\end{enumerate}
Hamilton's principle function is then completely separable as follows, $S(q_i, \alpha_i,t) = S_r(r) + S_\vartheta(\vartheta) + S_\varphi(\varphi) + S_t(t)$, or 
\begin{align} \label{eq:hamiltons-principle-function}
    S = \int \dd r &\sqrt{2m \left(E - m\Phi - mQ - \frac{g\rho}{m}\right) - \frac{L^2}{r^2}} \nonumber \\
    &\quad \quad + \int \dd \vartheta \sqrt{L^2 - \frac{L_z^2}{\sin^2 \vartheta}} + \int \dd \varphi L_z - Et.
\end{align}
We are interested in calculating orbits\footnote{We use the term "orbit" throughout this paper, knowing that "orbit" in the fuzzy regime entails a different nature from what is usually understood as "orbit" - an individual, spatial trajectory without extension. Since we will be only concerned with the TF regime in this paper, where we replace $Q$ by an effective classical velocity dispersion pressure, we can safely use the term "orbit" without confusion.} in a system described by (\ref{eq:hamiltons-principle-function}), i.e. we consider bound cases for which the total energy is negative. Hence, the motion in each of the coordinates will be periodic - libration in $r$ and $\vartheta$, and rotation in $\varphi$. We can thus introduce "action-angle variables", canonically conjugated variables familiar from galactic dynamics and celestial mechanics. The action variables $J_i$ (the new momenta) are defined as
\begin{align} \label{eq:actions}
    J_i = \frac{1}{2\pi} \oint p_i \dd q_i
\end{align}
with the (old) coordinates $q_i$ and their corresponding momenta $p_i$, and the line integrals are evaluated along a complete orbital period in the $(q_i,p_i)$-plane.
The associated new coordinates, the angle variables, are given by
\begin{align} \label{eq:angles}
   \theta_i = \frac{\partial W}{\partial J_i},~~~ \mbox{or}~~ \dot{\theta}_i = \frac{\partial \mathcal{H}}{\partial J_i} = \Omega_i,
\end{align}
where the angular frequencies $\Omega_i$ are constant functions of the $J_i$, and the angle variables depend linear on time, $\theta_i = \Omega_i t + \beta_i$ with constants $\beta_i$.
The set of variables $(J_i, \theta_i)$ with $i=r,\vartheta,\varphi$ defines an invariant 3-torus.
Combining then equation \eqref{eq:actions} with equations \eqref{eq:momentum-r}-\eqref{eq:momentum-phi}, we identify the action variables of the system as
\begin{align}
    J_\varphi &= \frac{1}{2\pi} \int_0^{2\pi} \dd \varphi L_z = L_z, \label{eq:action-phi} \\
    J_\vartheta &= \frac{1}{2\pi} \int_{\pi/2}^{\pi - \vartheta_\text{min}} \dd \vartheta \sqrt{L^2 - \frac{L_z^2}{\sin^2 \vartheta}}, \label{eq:action-theta} \\
    J_r &= \frac{1}{2\pi} \int_{r_\text{min}}^{r_\text{max}} \dd r \sqrt{2m \left(E - m\Phi - mQ - \frac{g\rho}{m} \right) - \frac{L^2}{r^2}} \label{eq:action-r},
\end{align}
with $\vartheta_\text{min}$ as the minimum value the polar angle occupies, and where $r_\text{min}$ and $r_\text{max}$ represent the pericenter and apocenter, respectively. Given all our symmetry assumptions, we see that $J_\varphi$ is trivial and $J_\vartheta$ can be readily calculated analytically, see e.g. \citet{binneytremaine}. The integral for $J_r$ is the most complicated one, and closed-form solutions are only possible for very few special cases, even in systems without quantum potential and SI, i.e. even if we abandon $Q$ and set $g = 0$. In fact, for our orbit calculations, we will disregard $Q$ altogether but keep the SI term with $g > 0$. This is justified, if we consider only the orbits within the TF cores of our halos, which is our focus here as the novel feature that we need to analyze, while the modelling of the halo envelope as an isothermal sphere presents nothing new in terms of its orbital structure\footnote{In a future paper, we will move beyond the current halo model, such that we can smoothly calculate orbits that can stretch throughout the entire halo -- core and envelope.}. 
So, in order to proceed with our calculation, we insert into (\ref{eq:action-r}) our expressions for $\rho$ and $\Phi$ from equations (\ref{eq:polytrope-density}) and (\ref{eq:polytrope-gravpot}), respectively, for the TF core. As a result, the integral cannot be solved analytically, however.
Therefore, we use a software package to calculate (all) the action variables numerically, and to this end we choose the Python code \texttt{gala}\footnote{\url{http://gala.adrian.pw/en/latest/}} \citep{gala}, which is a code to perform tasks in galactic and gravitational dynamics, such as orbit integrations, gravitational potential and force evaluations, or dynamical coordinate transformations. The functionality used for our purpose here follows the mathematical procedure described in \citet{sanders2014MNRAS.441.3284S}, i.e. it relies on the method of torus mapping developed by \citet{McGill_Binney_Torusmapping}. We provide the code with our density and gravitational potential, and a set of initial phase-space coordinates, $(r, \vartheta, \varphi)$ and $(v_r, v_\vartheta, v_\varphi)$, which are then evolved along an orbit for several time steps, up to the total integration time which we choose to be 2000 Myrs, using a leapfrog integrator. The obtained phase-space coordinates are subsequently transformed to action-angle variables via a generating function, which maps the "toy torus" of one of two "toy potentials" - in our case the isochrone potential - into the desired "true torus" that describes the orbit of interest\footnote{The applied "toy potentials" are divided into two categories: a triaxial harmonic oscillator is used for box orbits, while the isochrone potential is used for loop orbits. Since the spherical $(n=1)$-polytropic potential considered here is part of the latter family, the isochrone potential is our toy potential of choice.}. 

\begin{figure*}
    \centering
    \includegraphics[width=0.33\textwidth]{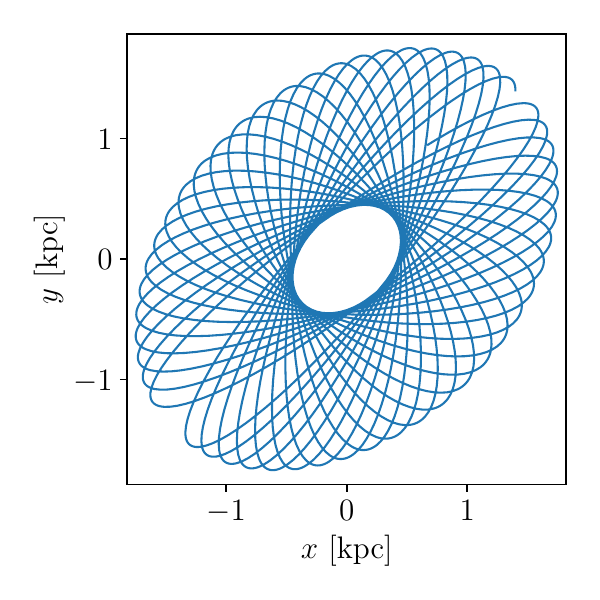}
    \includegraphics[width=0.33\textwidth]{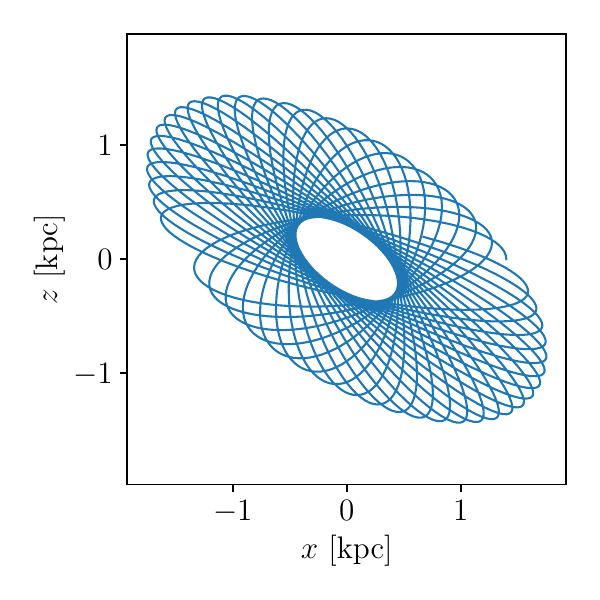}
    \includegraphics[width=0.33\textwidth]{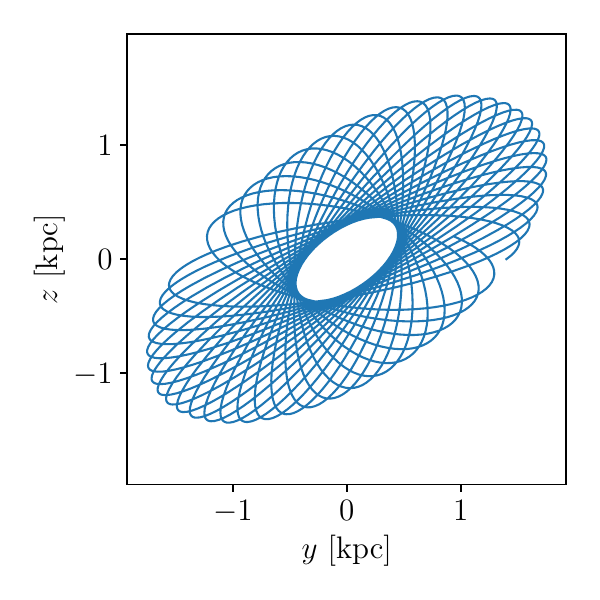}
    \caption{\textbf{Dark matter orbits within the TF core} with a core radius of $R_\text{TF} = 4$ kpc and SFDM parameters, $m = 10^{-22}$ eV/$c^2$ and $g = 2.426 \cdot 10^{-61}$ eV cm$^3$. The origin of the Cartesian coordinate system is situated in the center of the halo core. With arbitrarily chosen initial conditions of $(r, \vartheta, \varphi) = (2$ kpc$, 0$ rad$, 0.78$ rad$)$ for the positions and $(v_r, v_\vartheta, v_\varphi) = (30$ km s$^{-1}, 15$ km rad kpc$^{-1}$ s$^{-1}, 15$ km rad kpc$^{-1}$ s$^{-1})$ for the velocities, the orbits are integrated over 2000 Myrs, resulting in loop orbits which are typical for spherical and axisymmetric systems.}
    \label{fig:orbits}
\end{figure*}

\begin{figure*}
    \centering
    \includegraphics[width=\textwidth]{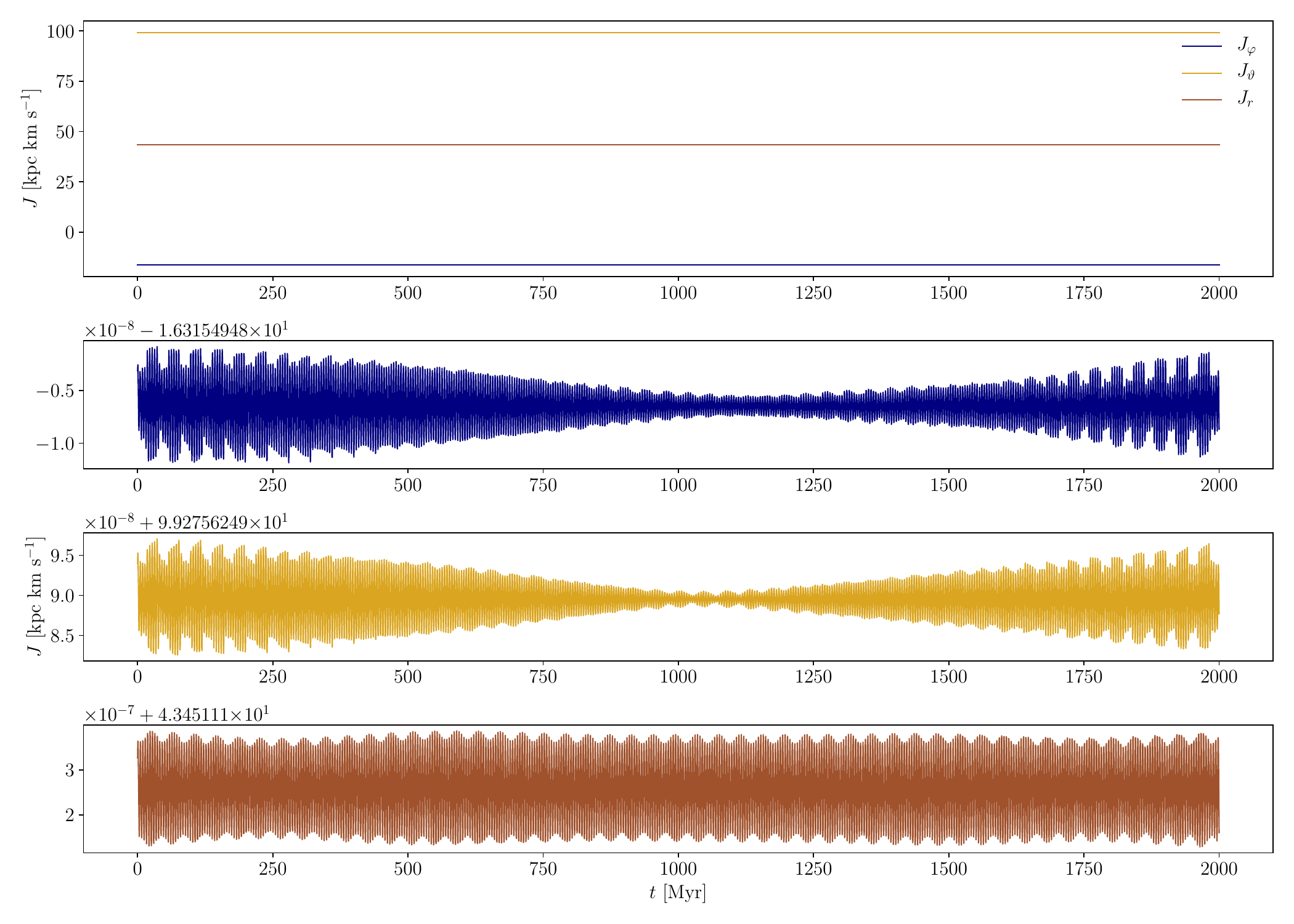}
    \caption{\textbf{Action variables \eqref{eq:action-phi}, \eqref{eq:action-theta} and \eqref{eq:action-r} over time, generated by the orbits and initial conditions of Figure \ref{fig:orbits}.} The \textit{top} panel demonstrates that all three action variables are basically constant over time, as expected. The \textit{bottom} three panels provide zoom-in versions for each action variable, where the graphs are colored according to the legend in the top panel. Minor oscillations are seen in the range of $10^{-8}$ to $10^{-7}$, as a result of numerical inaccuracies.}
    \label{fig:orbit_actions}
\end{figure*}

In Figure \ref{fig:orbits}, we show example orbits that result upon such a procedure. Apart from (\ref{eq:polytrope-density}) and (\ref{eq:polytrope-gravpot}), we have to set the TF core radius which depends upon $g/m^2$ - see (\ref{eq:tfradius}) - and we choose $R_\text{TF} = 4$ kpc. However, this does not yet fix $g/m$ in (\ref{eq:action-r}). Therefore, we have to specify a particle mass $m$, too. We choose $m = 10^{-22}$ eV/$c^2$ which, together with the TF radius, yields the associated value for the SI coupling strength,
$g = 2.426 \cdot 10^{-61}$ eV cm$^3$.
Furthermore, we choose spherical initial phase-space coordinates, as follows: $(r,\vartheta,\varphi) = (2$ kpc$, 0$ rad$, 0.78$ rad$)$, $(v_r,v_\vartheta,v_\varphi) = (30$ km s$^{-1}, 15$ km rad kpc$^{-1}$ s$^{-1}, 15$ km rad kpc$^{-1}$ s$^{-1})$.
In Figure \ref{fig:orbit_actions}, we show the action variables during the entire run up to $2000$ Myrs, obtained from \texttt{gala}, where the bottom three panels show zoomed-in versions of the individual actions from the top panel.
The top panel clearly shows the expected constancy of the action variables. Minor oscillations of order $10^{-8} - 10^{-7}$ can be seen in the zoom-in of the bottom panels; they stem most likely from numerical inaccuracies.

We also tested other initial conditions, or halo parameters. Generally, we find that the orbital eccentricity depends highly on the initial conditions, with higher velocities usually leading to more eccentric orbits. Independent of the initial conditions, the orbits always create rosettes, typical for loop orbits in spherical systems. We also find that the orbital structure is little affected by the SFDM parameters; e.g. a choice\footnote{This choice is more or less arbitrary; the fiducial model of \cite*{Li_2017} had the same particle mass, but a smaller $R_\text{TF}$.} of $m = 8\cdot 10^{-21}$ eV/$c^2$ and $g = 1.552\cdot 10^{-57}$ eV cm$^3$, using the same TF radius, will not visibly affect the orbits. This is not too surprising, given the small numbers involved. A detailed study of the impact of SFDM parameters on orbital structure is beyond the scope of this work, but will be presented elsewhere.  

In any case, we emphasize that the content of this section establishes the use and form of action-angle variables, appropriate for the novel dynamics presented by the TF halo \textit{core}, whose density and gravitational potential has been described above. Thereby, we are able to calculate orbits within that core. On the other hand, as described in the previous section, the halo envelope is modelled as an effective isothermal sphere, justified as a coarse-grained model which averages over the genuine SFDM dynamics on small scales. In this sense, we can pretend that the halo \textit{envelope} is just as "classical" as any other CDM halo envelope, with an orbital structure similar to any other system described by an isothermal sphere.

More precisely, we stress that the TF regime with its neglect of $Q$, or rather its replacement by an effective classical term, is conceptually simpler than the other limit case of FDM. Consequently, the fuzziness and non-locality introduced by $Q$ is neglected, or smoothed over, and we can think of orbits in our SFDM-TF halo models here as very close analogues to collisionless CDM particle orbits. Therefore, we can use adiabatic invariants, evaluated at any given orbital location, the same way as we would for CDM.
We turn to the adiabatic contraction (AC) routines in the next section.

\section{Adiabatic contraction} \label{sec:adiabatic-contraction}

\subsection{SFDM-TF plus baryons}
\label{sec:AC1}

In the last section, we have laid the foundation of calculating orbits within SFDM-TF halos by studying the novel dynamics presented by their cores, while the envelopes are modelled akin to CDM. Orbits constitute an implicit building block of the AC framework with which the dynamic response of DM due to baryons in galactic halos of interest is calculated, because adiabatic invariants are evaluated as a function of orbital coordinates.

In this subsection, we apply the "Blumenthal method" \citep{blumenthal1986} to calculate for the first time adiabatically modified SFDM-TF halo profiles due to the presence of baryons. In the next subsection, we use a modified version by \cite{Gnedin_2004} ("Gnedin method") in order to study whether the use of different AC routines has an impact on our results. We find that this is the case, especially for large TF cores, $R_\text{TF} \gtrsim 1$ kpc. Finally, in Section \ref{sec:53}, we compare our results to observational data of dwarf galaxies and to CDM, whereas Section \ref{sec:54} includes a brief comparison to FDM. 

In the original Blumenthal method, the basic assumptions are spherical symmetry, circular orbits and angular momentum conservation, such that the product $r M(r)$, where $M(r)$ denotes the enclosed mass at radius $r$, remains constant for any orbit evaluated at a radius $r$. This is related to the adiabatic invariants, determined from equations (\ref{eq:action-phi})-(\ref{eq:action-r}).
The final dark matter mass distribution $M_\text{dm}$ is calculated by
\begin{align} \label{eq:adiabatic_contraction}
    r_f \left[ M_\text{dm}(r_f) + M_b (r_f) \right] = r_i M_i (r_i) = r_i \frac{M_\text{dm}(r_f)}{1 - f},
\end{align}
with an initial total mass profile $M_i$ at the initial radius $r_i$, a final baryon mass profile $M_b$ at the final radius $r_f$, and a baryon fraction $f$ with $0 < f < 1$ (we follow common nomenclature; $f$ shall not be confused with the subscript "$f$" in $r_f$). 
The initial total mass profile requires a combined input profile for the DM and the baryons, while the desired output of interest concerns the DM mass profile. 

We solve equation \eqref{eq:adiabatic_contraction} in an iterative manner, using a Python code whose core concept was taken from the implementation of \cite{Freese2009}, appropriately modified and extended to fit our needs. First, we convinced ourselves that our code works correctly and accurately by reproducing the results of \cite{blumenthal1986}, namely their Figures 1 and 2 for the same choice of parameters and profiles.

In our study of AC applied to SFDM-TF, we have to specify input profiles and parameters, which we choose as follows:
\begin{itemize}
    \item \textit{Initial profile:} Here, we need a combined input profile for the DM and baryonic components. We use the core-envelope halo structure appropriate for SFDM-TF, as discussed in Section \ref{sec3}. Since \cite{Dawoodbhoy_2021} showed that realistic halo profiles require a value of $\chi$ close to one, we simply pick $\chi = 1$ in all of our runs. In studying the impact of different core radii, we vary $R_\text{TF}$ in the range $[0.1,4]$ kpc. For simplicity, we adopt the \textit{same} core-envelope profile for the baryonic component. This is not a severe restriction; typically the same simplification is used for CDM halos, where an initial NFW profile \citep{nfw} is used for both DM and baryons, because the final results are not much affected, see e.g. \cite{Freese2009}.
    \item \textit{Final baryon profile:} We choose a Hernquist profile \citep{hernquist1990},
    \begin{align}\label{eq:Hernquist}
        \rho(r) = \frac{\rho_s}{\frac{r}{b} \left( 1 + \frac{r}{b}\right)^3},
    \end{align}
      motivated by its prevalence in similar studies on dwarf galaxies. We tested three representative scale lengths: $b=0.03$ describes a small spheroidal, $b=0.07$ is an intermediate case and $b=0.2$ accounts for a more extended galaxy.    
    \item \textit{Baryon fraction:} We are interested in dwarf galaxies with a dynamical mass of roughly $\sim 10^{10}-10^{11}~M_{\odot}$, so we would like to make an informed choice concerning the baryon fraction $f$, which is a free parameter in the adiabatic invariant of equation (\ref{eq:adiabatic_contraction}). Unfortunately, there seems to be considerable uncertainty and variability of $f$ from observations. In order to obtain a rough estimate for $f$, we follow \citet{Brook+diCintio_2015}, who studied the correlation of stellar-to-halo-mass for 40 Local Group dwarf galaxies. On the basis of \cite*{Posti_2019} we use this ratio to determine the baryon fraction to be in the range of $\sim 0.1 - 0.2$. Therefore, we adopt a low-baryon case with $f = 0.07$, as well as cases with $0.1$ and $0.13$, respectively. These values coincide with those that \citet{Mocz2020} find in their simulated FDM halos, to which we will get later in this section. A fourth case further represents the cosmic mean baryon fraction of $0.157$, based on the cosmology from \citet{Planck_2014}.
    \item \textit{Halo parameters:} Finally, we need to specify the halo virial mass $M_h$ and virial radius $R_h$. Using equation \eqref{eq:velocity-dispersion} and assuming isothermality, these are related to the effective velocity dispersion according to
    \begin{equation}  
        \sigma^2 = \frac{GM_{h}}{R_{h}} = \text{constant}.
    \end{equation}
   The constancy of the velocity dispersion effectively guarantees isothermality, which is appropriate for the halo envelope, see Section \ref{subsec:core-envelope}. However, as emphasized in \cite{Dawoodbhoy_2021}, we can vary the input parameters and either specify $\{R_\text{TF}, \sigma^2\}$, or $\{R_\text{TF}, \rho_0\}$.  
\end{itemize}

\begin{figure*}
    \centering
    \includegraphics[width=\textwidth]{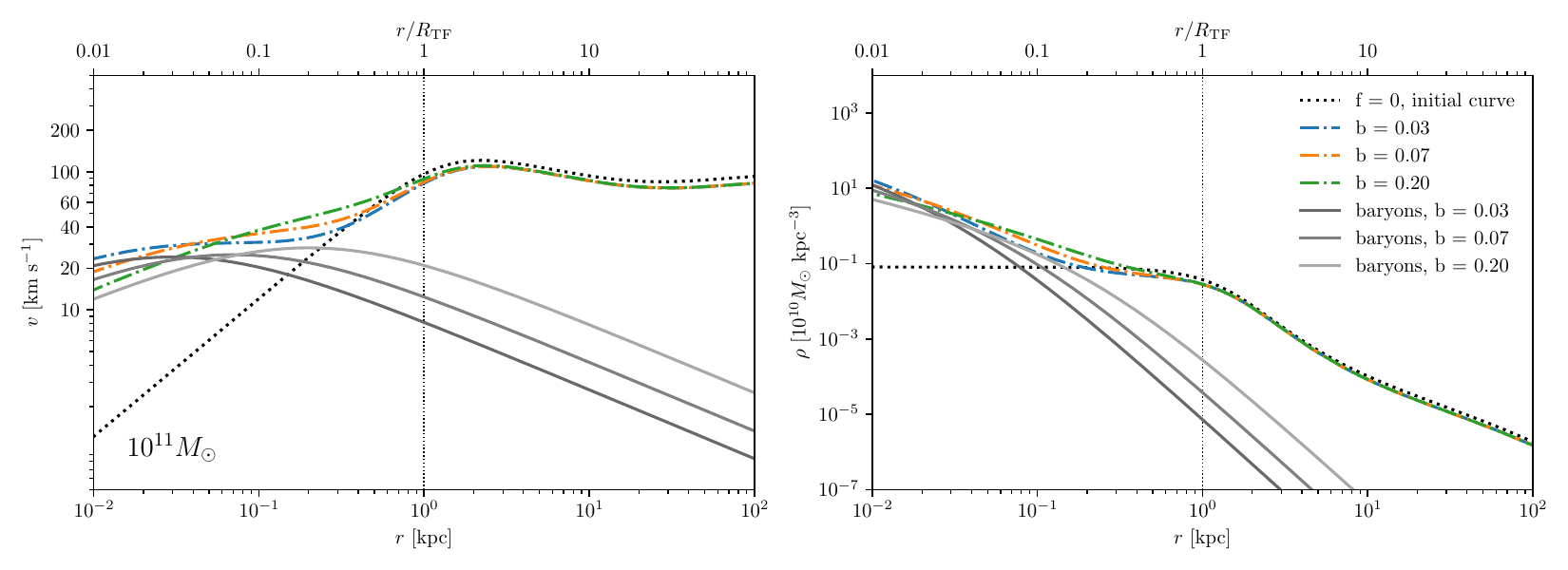}
    \caption{\textbf{Adiabatic contraction of a SFDM-TF halo} with a halo mass of $M_h=10^{11} ~M_{\odot}$, halo radius of $R_h = 96$ kpc and a core radius of $R_\text{TF} = 1$ kpc. In this figure, we fix the baryon fraction to $f = 0.1$, but vary the scale length parameter $b=0.03, 0.07, 0.2$ (dash-dotted lines in color). The initial pure-DM profile ($f=0$) is plotted as the dotted curve. The baryon component of the final profiles of the same $b$ are plotted as solid grey curves. \textit{Left panel:} Circular velocity profiles. \textit{Right panel:} Density profiles.  }
    \label{fig:different_scalefactor}
 \end{figure*}

As a first plot, we show in Figure \ref{fig:different_scalefactor} adiabatically modified SFDM-TF-plus-baryon velocity and density profiles for a typical "cusp-core halo" of mass $M_h = 10^{11} \, M_\odot$ and radius $R_h = 96$ kpc. We fix the core radius to $R_\text{TF} = 1$ kpc. The profiles probe several baryonic scale length parameters $b$ (see (\ref{eq:Hernquist})), as indicated in the legend in the right panel, but the baryon fraction is fixed to $f=0.1$. The initial pure-DM profile ($f=0$) is plotted as the dotted curves in each panel. The baryon component of the final profiles for the same $b$ are shown with solid grey curves. We can see that most changes of the final post-AC profiles are in the inner parts, while the envelope is barely affected. As expected, the initially cored central DM density is increased by orders of magnitude, here $\sim 100$, with enhancements throughout the entire core region. The baryon profiles are more extended, the larger the value of $b$, i.e. the impact on the DM density is correspondingly bigger, at least down to scales of roughly $5\cdot 10^{-2}~ R_\text{TF}$. In a real in-depth comparison with data, the scale length $b$ should be ideally determined from the data. However, our choice of baryon profile is arbitrary, and other choices may fit better in a case-by-case comparison of different galaxies. Given this freedom and the fact that we are more interested in the impact of the SFDM core radii and baryon fractions, we will choose a fixed $b=0.07$ for the remaining plots in this paper, however.

\begin{figure*}
    \centering
    \includegraphics[width=\textwidth]{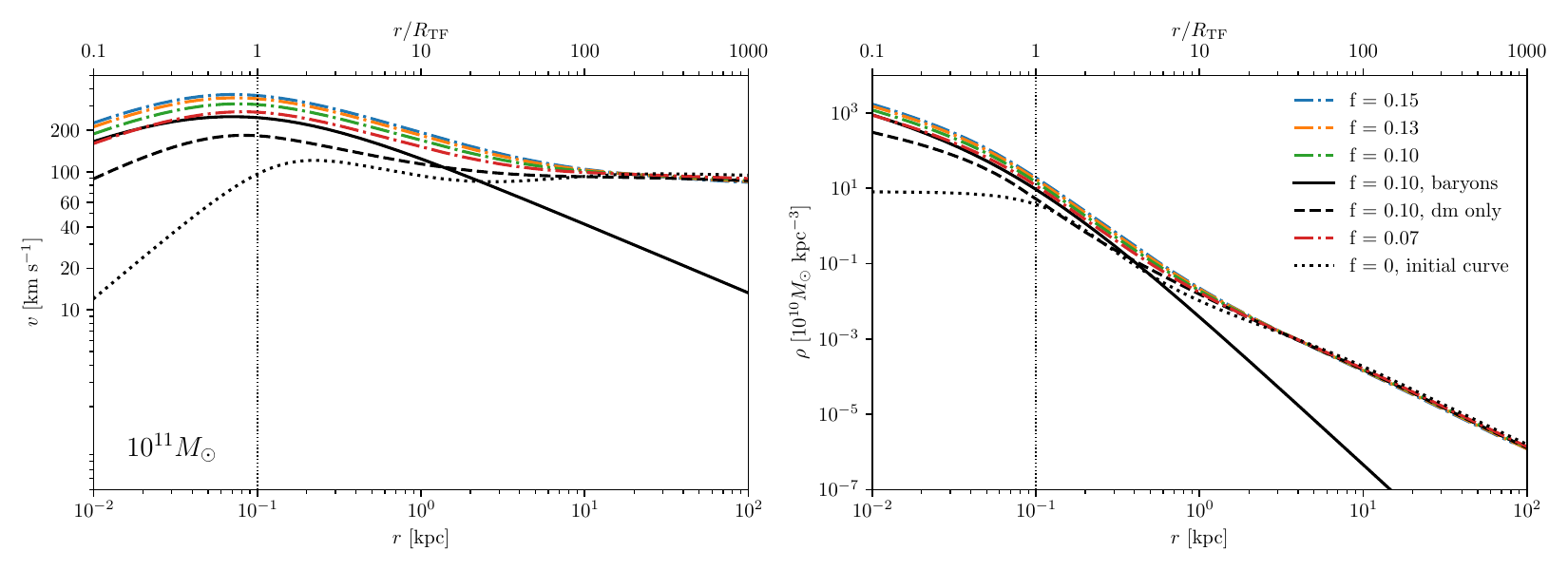}
    \includegraphics[width=\textwidth]{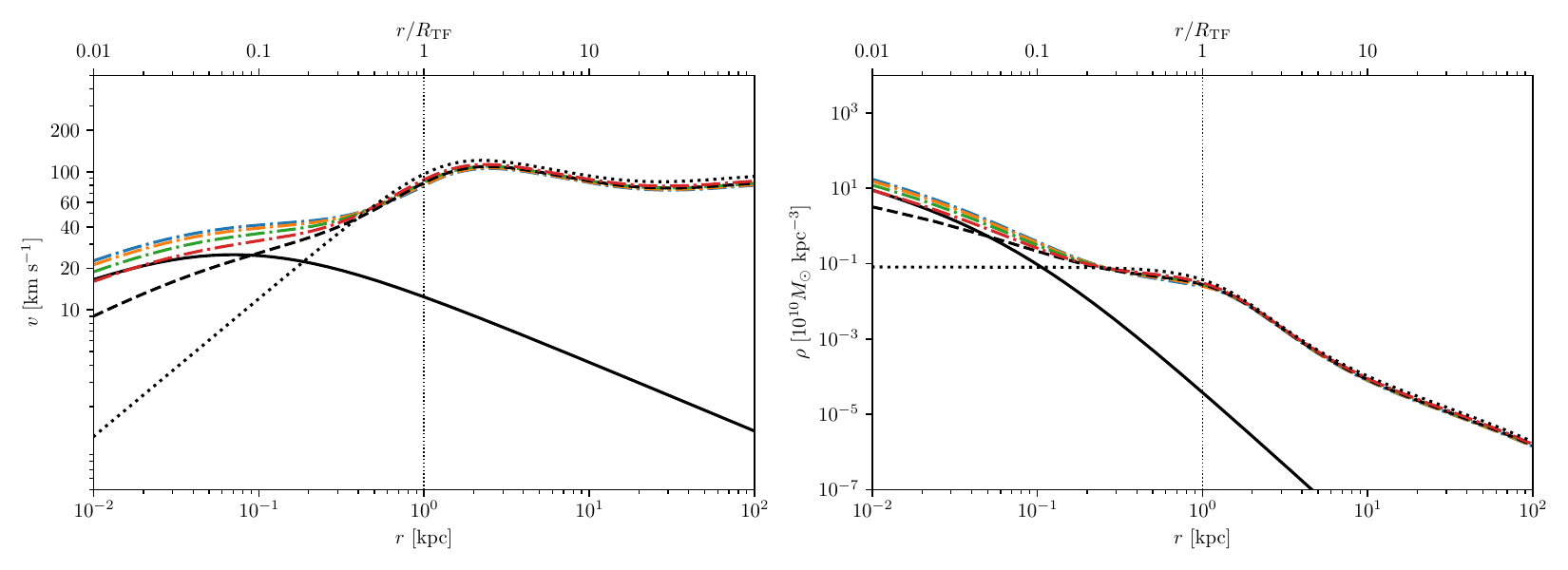}
    \includegraphics[width=\textwidth]{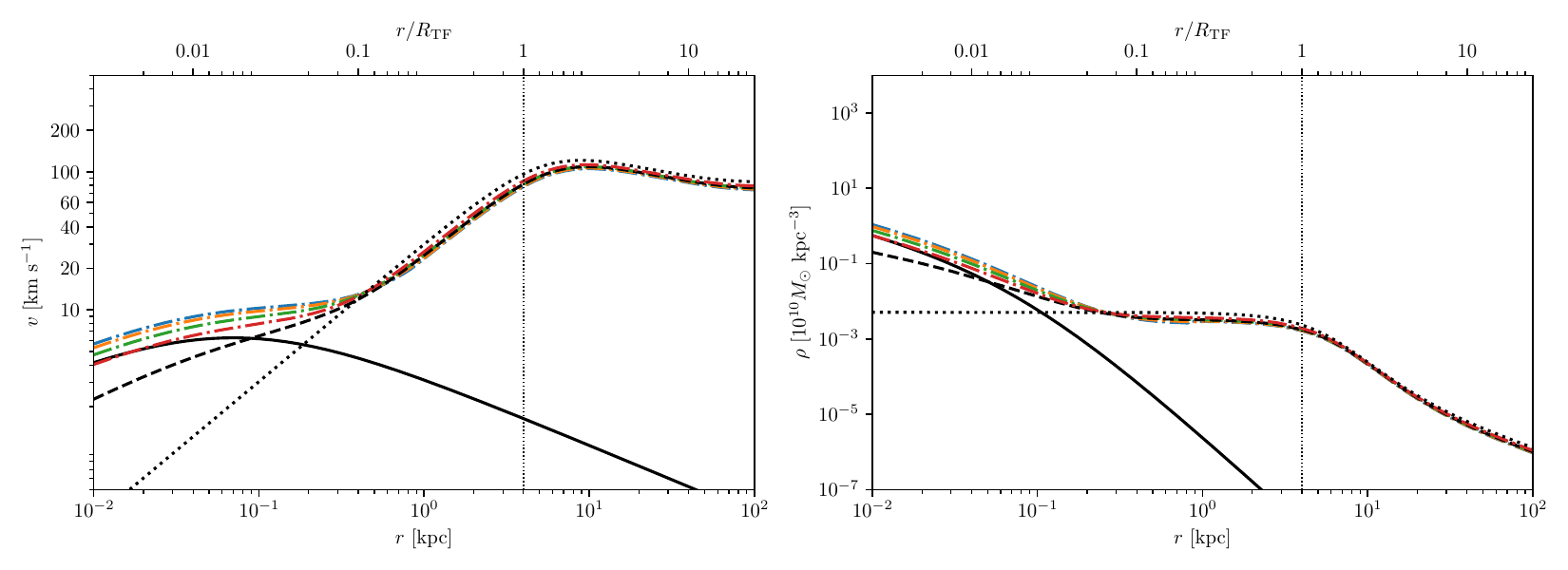}
    \caption{\textbf{Adiabatic contraction of a SFDM-TF halo} for a halo mass of $M_h = 10^{11} \, M_\odot$, halo radius of $R_h = 96$ kpc and core radii $R_\text{TF} = 0.1, 1 , 4$ kpc (top to bottom). Each panel probes several baryon fractions, as indicated in the legend in the top right panel with the overall profiles in various colors according to the legend. The initial pure-DM ($f=0$) profile is shown as the dotted curve. The $f=0.1$ case is shown as the total profile (green dashed), and separately as its DM-component (black dashed) and  baryon-component (black solid), respectively. \textit{Left panels} show circular velocity profiles, \textit{right panels} show the corresponding density profiles.}    
    \label{fig:different_core_radii}
\end{figure*}

In Figure \ref{fig:different_core_radii}, we show now adiabatically modified SFDM-TF velocity and density profiles for the same halo as in Fig. \ref{fig:different_scalefactor}, but for different core radii $R_\text{TF} = 0.1,1,4$ kpc (top to bottom panels). Each case probes several baryon fractions, as indicated in the legend in the top right panel. The initial pure-DM profile ($f=0$) is plotted as the dotted curves in each panel, and serves again as a comparison.  

In general, we see the expected overall trend that AC due to baryons affects the outer halo only barely or not at all, which is reasonable because DM dominates in the outer parts of halos. On the other hand, the more central halo parts, where baryons dominate, change dramatically: again, as a result of AC the central density increases by factors of $\sim 100$, while central velocities increase by factors of $\sim 10-100$.
Also, we can see that higher baryon fractions correlate with higher densities and velocities in the central parts, which is an expected outcome. 

However, the more interesting dependency for our studies concerns the impact of the initial ("primordial") SFDM-TF core, as follows.  
Small (sub-kpc) cores have higher initial core densities, compared to large ($\sim$ kpc) cores, according to
\begin{equation}\label{eq:densscale}
\rho_c \propto \frac{M_c}{R_\text{TF}^3} \propto M_c \propto M_h^{2/3},
\end{equation}
see equation (104) in \cite{Dawoodbhoy_2021}. (This scaling between core mass $M_c$ and total halo mass $M_h$ holds under the assumption of isothermality.)
Therefore, once the DM gets compressed through AC, this trend is enhanced, and we can see in Figure \ref{fig:different_core_radii} that models with small cores have higher central densities. Hence, small initial SFDM cores post-AC can lead to very high central densities and velocities, in potential or actual conflict with galaxy data. For instance, Fig. \ref{fig:different_core_radii} shows that the case with $R_\text{TF} = 0.1$ kpc leads to values of $\gtrsim 200$ km s$^{-1}$ at $r = 0.1 R_\text{TF}$, whereas a core radius of $R_\text{TF} = 4$ kpc yields $\gtrsim 10$ km s$^{-1}$ at $r = 0.1 R_\text{TF}$. Of course, in terms of physical spatial scales in units of kpc, the effect looks more dramatic. It is observationally very challenging to measure the presence of a small core, let alone its central impact, but the figure shows clearly that velocity data as close as possible to the center of dwarf galaxies is highly warranted when it comes to confirm or to rule out SFDM models (generally all DM models with potentially small cores, in fact).  This last point of issue is also of relevance to the next subsection.

\subsection{Modified adiabatic contraction}
\label{appendix:modified-AC}

In Section \ref{sec:AC1}, we used the adiabatic contraction model from \cite{blumenthal1986} with its assumption of purely circular orbits. In reality, however, this is just an approximation. Indeed, simulations of CDM halos that are not limited to simplifying assumptions, like spherically-symmetric galactic potentials, reveal highly eccentric orbits \citep{Ghigna_et_al_DMorbits}, or centrophilic orbits whose structure remains relatively stable even after adiabatic contraction  \citep{Valluri_etal_2010, Valluri_etal_2012}. It is fair to expect a similar diversity in general SFDM halos which lack symmetries. The study of such general potentials is outside the scope of this paper, but we can still check for some modification of the Blumenthal method that goes beyond the assumption of circular orbits, including the presence of eccentric orbits in an "effective" way. After all, even the loop orbit examples plotted in Figure \ref{fig:orbits} exemplify some eccentricity. Moreover, we deem it worthwhile to have another comparison model concerning the AC routine.
Therefore, we will employ the method by \cite{Gnedin_2004}, which was motivated by CDM simulation results, and which modified the original Blumenthal method in order to incorporate eccentric orbits, as follows. Instead of using the adiabatic constant $r M (r)$, \cite{Gnedin_2004} use $r M (\Bar{r})$, where $\Bar{r}$ denotes the orbit-averaged radius. It is defined as
\begin{align}
    \Bar{r} = r_\text{vir} A (r/r_\text{vir})^w
\end{align}
with halo-to-halo variations of the parameters $A$ and $w$, and the halo virial radius $r_\text{vir}$. This choice is somewhat empirically motivated: upon testing different options and comparing to simulations, the quantity $r M(\Bar{r})$ gave the best results. 

We repeat our AC calculations from the previous subsection, but now apply this modified AC provided by \citet{Gnedin_2004}. Of course, we have no SFDM simulations to gauge and optimize this approach in the first place, so we merely stick to the "mean values" applied by \citet{Gnedin_2004}, which are $A \approx 0.85 \pm 0.05$ and $w \approx 0.8 \pm 0.02$, respectively. The Blumenthal invariant is recovered for $A=1=w$, so these "mean values" found originally by \citet{Gnedin_2004} are not very different from $1$. Yet, these AC parameters are subject to various conditions, e.g. \citet{Duffy2010} studied best-fit values of these parameters at redshifts of $z = 0$ and $z = 2$. Depending upon the incorporated baryon feedback (none, weak or strong stellar feedback), their values range between $A \in [0.1, 1]$ and $w \in [0.111, 0.556]$.

The Gnedin modification of AC not only takes into account eccentric DM orbits, but it also resolves the general problem with the Blumenthal method which tends to predict a higher overall density contraction in the inner parts of halos, $r / r_\text{vir} \lesssim 0.1$, compared to cosmological simulations. In fact, \citet{Gnedin_2004} find that their modified AC model not only results in deviations of only $\lesssim 10$\% compared to such simulations, but that it also avoids systematic "overpredictions" of the resulting central densities, in contrast to the Blumenthal method.

\begin{figure*}
    \centering
    \includegraphics[width=\textwidth]{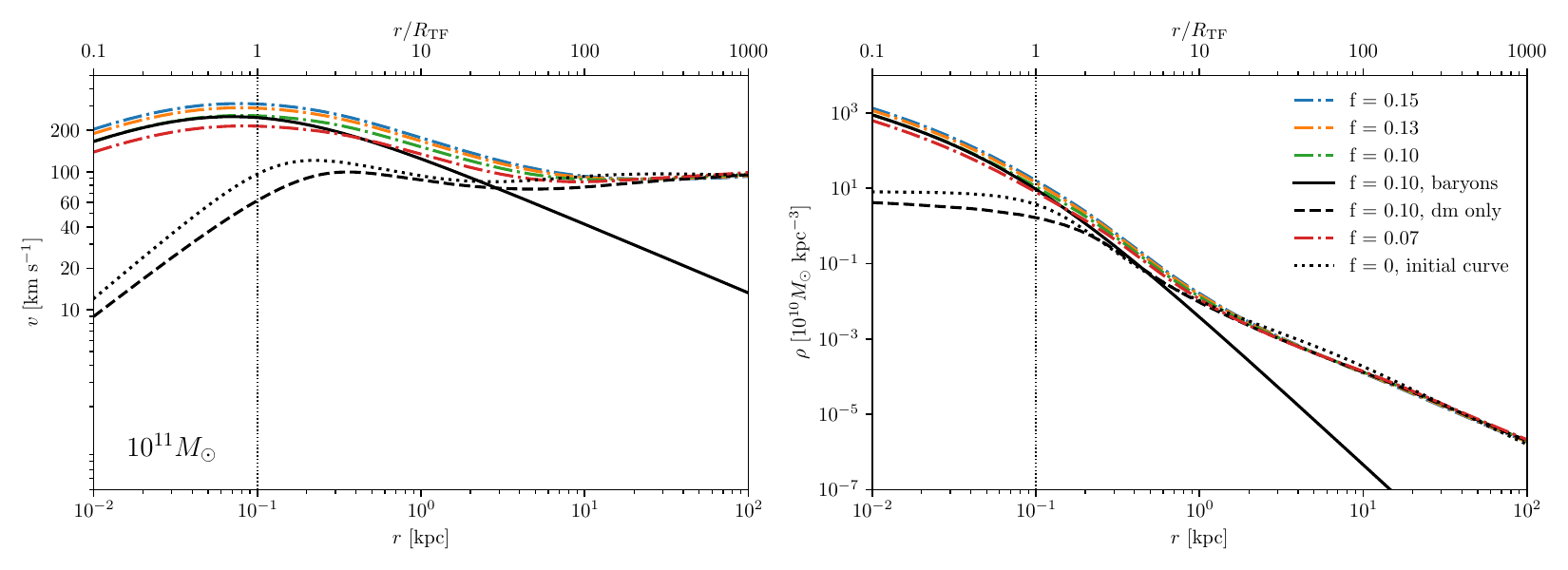}
    \includegraphics[width=\textwidth]{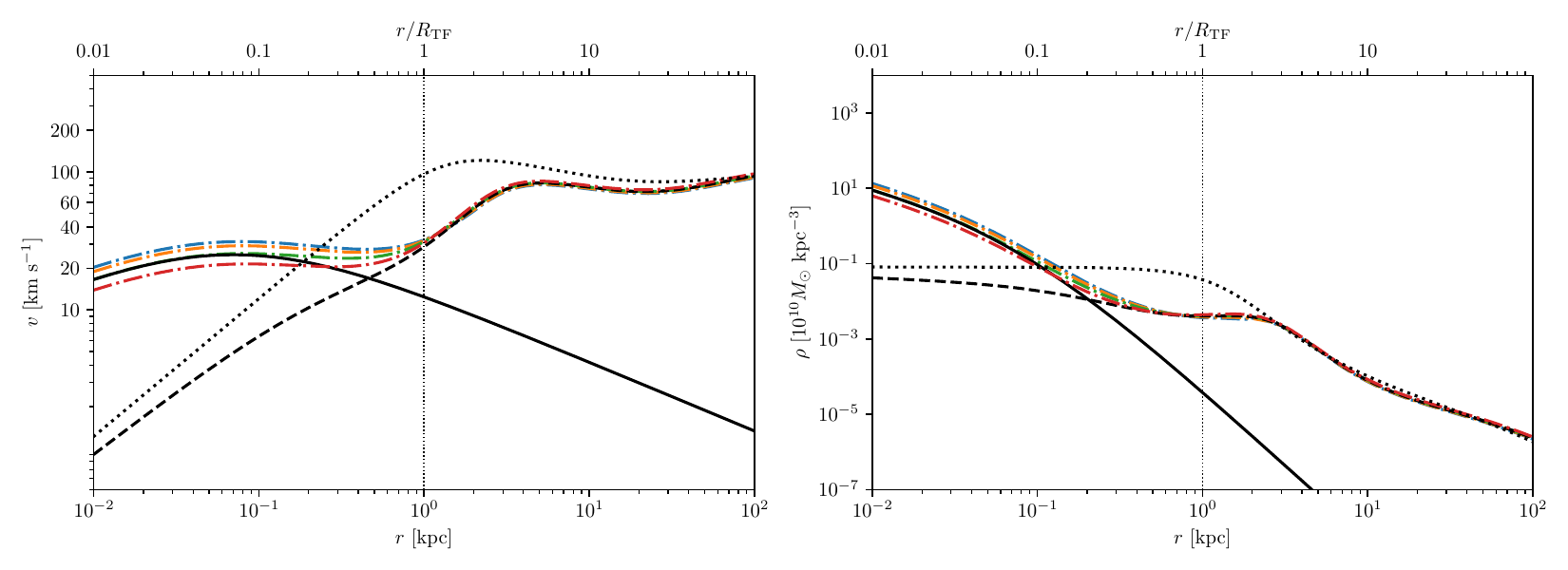}
    \includegraphics[width=\textwidth]{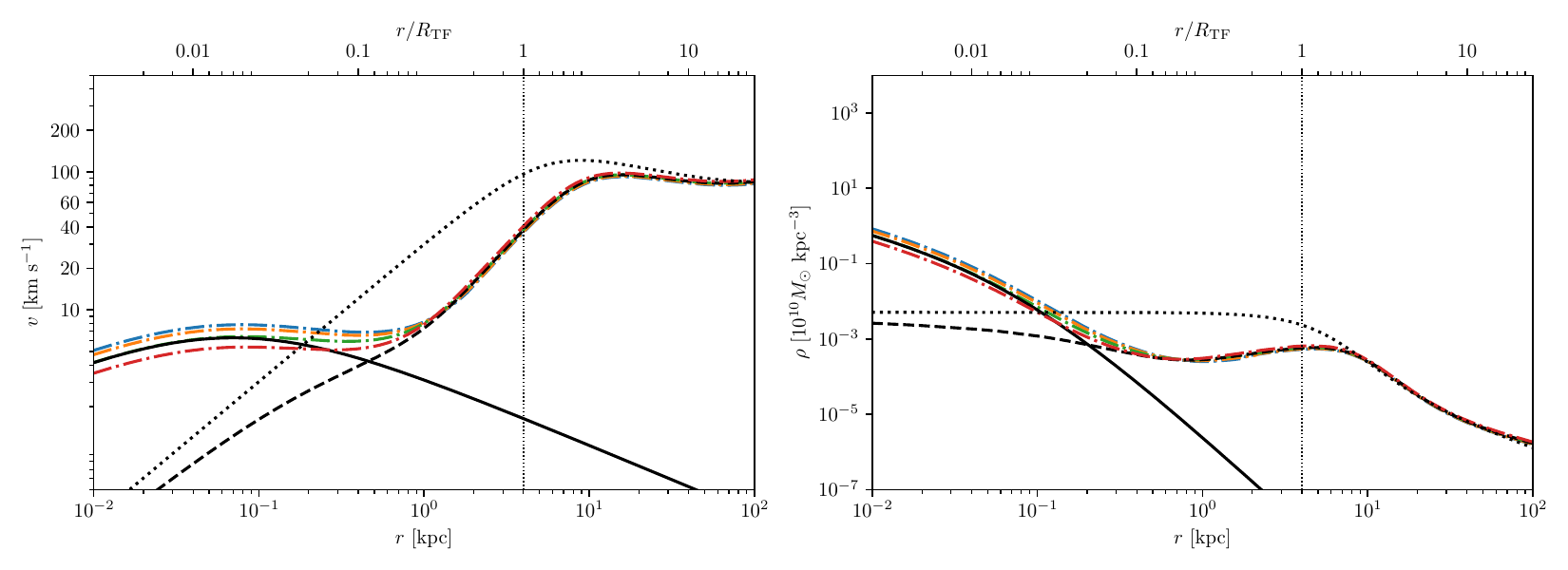}
    \caption{\textbf{Adiabatic contraction of a SFDM-TF halo using the method by} \citet{Gnedin_2004}. The halo parameters are identical to those in Figure \ref{fig:different_core_radii}, i.e. $M_h = 10^{11}~M_{\odot}$, $R_h = 96$ kpc, with different core radii of $R_\text{TF} = 0.1, 1, 4$ kpc, respectively, from top to bottom. See also caption of Figure \ref{fig:different_core_radii}.}
    \label{fig:different_core_radii_gnedin}
\end{figure*}

Figure \ref{fig:different_core_radii_gnedin} shows velocity and density profiles for the same choice of halo and baryon parameters as in Figure \ref{fig:different_core_radii} before, but now using our modified AC procedure according to \citet{Gnedin_2004}. We broadly find similar results between the Blumenthal and Gnedin AC approach, in the sense that smaller initial TF cores lead to higher velocities and densities within these cores post-AC, in potential conflict with data of dwarf galaxies. In fact, for small $R_\text{TF}$, the two approaches give very similar results in terms of overall profile slopes, although we confirm the general decrease in density by a factor of roughly $2$ within the core radius $R_\text{TF}$, as a result of the adjustment for eccentric orbits. This difference in density also leads to a sharper decline in velocity at these radii. 

However, by the same token we observe that the reduction in density in the Gnedin approach causes a significant plateau or even "dip" in the density between around $r=(0.1-1)~R_\text{TF}$, which is more pronounced, the larger $R_\text{TF}$. A similar feature and trend can be seen in the velocity profiles. This feature occurs despite the fact that the "mean values" of $A$ and $w$ are in each case close to one. We checked to convince ourselves that smaller values of $A$ and $w$, compared to the "mean values" picked for the plots, lead to an exaggeration of this effect, causing a bigger discrepancy between the Blumenthal and Gnedin approaches. Again, this discrepancy is more critical, the larger the halo core. Furthermore, we checked the procedure with a reduced baryon density, and find that this leads to an overall reduced final profile, instead of only a "dip"-feature around $R_\text{TF}$.

This is an interesting result, although it points to a complication, because it confirms that the interpretation of observational data will be subject to such modelling details as the use of different AC invariants. Since the properties of the inner core and the transitional region to the outer envelope depend upon the DM parameters, high-quality observational data is required to disentangle features caused by DM microphysics from those due to the modelling of DM dynamics (here AC).

\subsection{Comparison to data and CDM}
\label{sec:53}
 
\begin{figure*}
    \centering
    \includegraphics[width=0.49\textwidth]{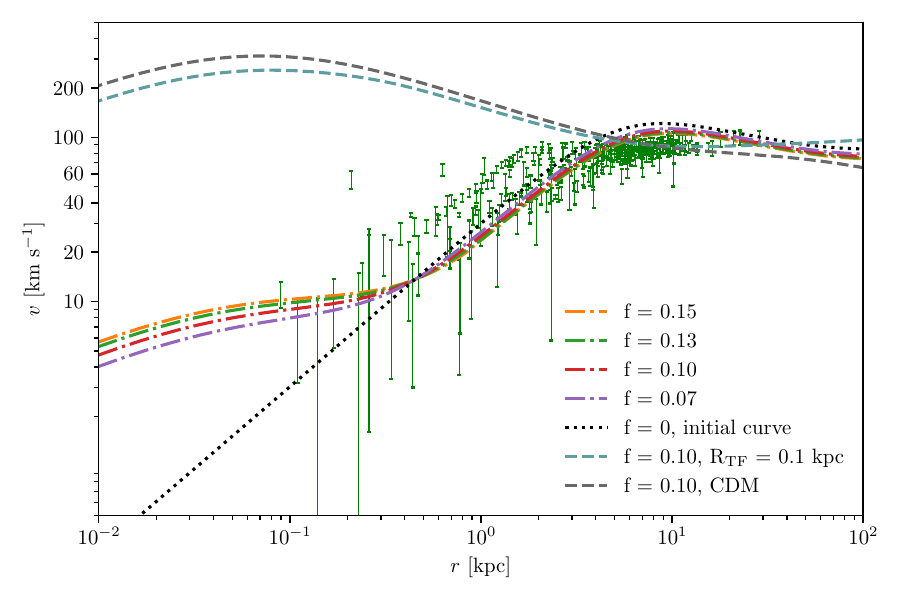}
    \includegraphics[width=0.49\textwidth]{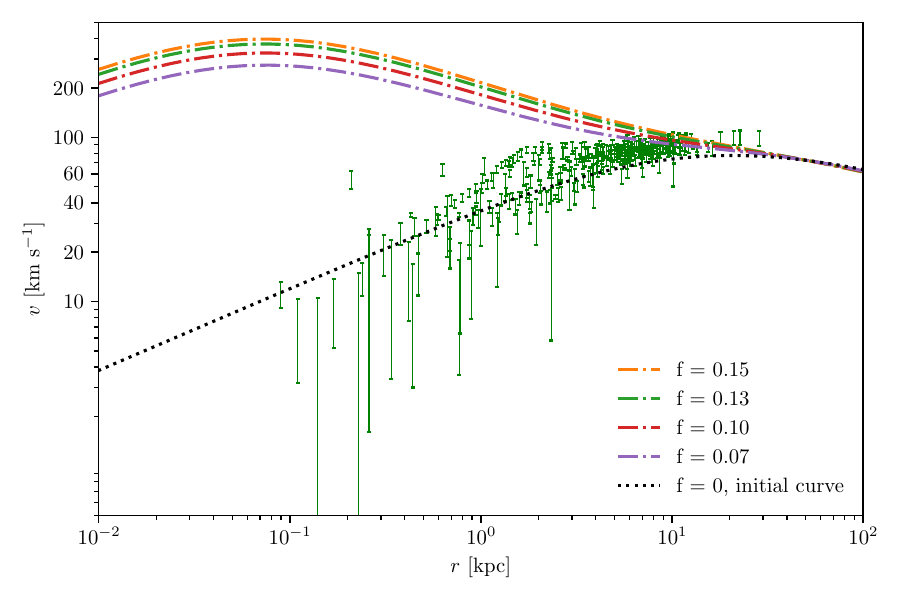}
    \includegraphics[width=0.49\textwidth]{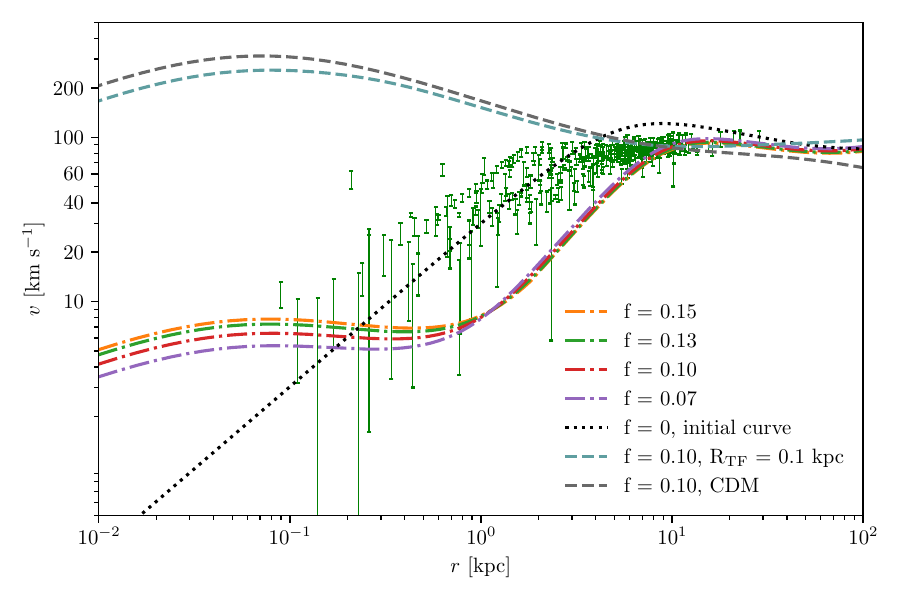}
    \includegraphics[width=0.49\textwidth]{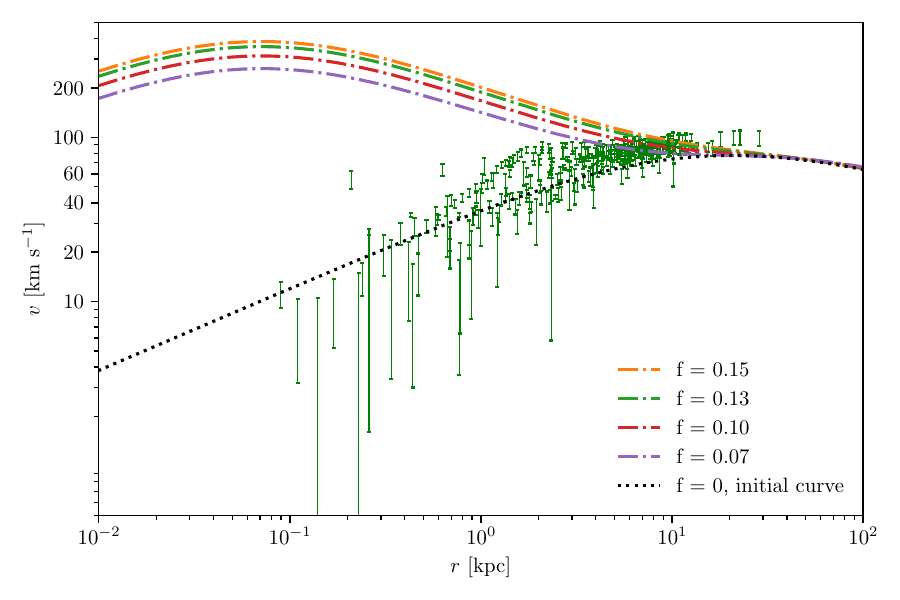}
    \caption{\textbf{Adiabatically modified DM velocity profiles in comparison to observational dwarf galaxy rotation curve data from} \citet{Lellietal}. The data sample, shown here in thin green bars, consists of 18 dwarf galaxies with asymptotic circular velocities between $80 - 100$ km s$^{-1}$. \textit{Top} panels show models calculated with the Blumenthal AC method, \textit{bottom} panels show results using the Gnedin AC method. \textit{Left} panels show SFDM and CDM models, while \textit{right} panels show CDM only. 
    For comparison, the respective initial DM profile ($f=0$) is shown as black dotted curve in each panel, for comparison (in the left panels this curve refers to SFDM-TF). Each model depicts the profiles for a halo with $M_h = 10^{11}~M_{\odot}$, $R_h = 96$ kpc throughout all panels, with different baryon fractions according to the legends. 
    The CDM model is parameterized with a NFW profile with $c=10$.
    Left panels: The colored, dash-dotted curves show SFDM-TF models with different baryon fraction, but fixed core radius $R_\text{TF} = 4$ kpc. For comparison, we also show a SFDM-TF model with small core ($R_\text{TF} = 0.1$ kpc; green dashed curve) and the CDM model (blue dashed curve), each with $f=0.1$.
    } 
    \label{fig:Datacomparison}
\end{figure*}

In this subsection, we compare our results with data and CDM. 
It is customary to point to the mere cusp-core difference between CDM (predicting cusps) and non-CDM models (predicting cores), but there are many nuances that all require care. As discussed in the Introduction, the size of cores in non-CDM models such as SFDM is subject to the arbitrary choice of DM particle parameters.  Ideally, observations would provide bounds on the core size in order to confirm or rule out DM models, and this happens in many cases, e.g. when core size is compared to the minimum size of galaxies in the Local Group, yielding upper bounds e.g. on $\lb$ \citep{Nadler_etal2021}, relevant to FDM. However, as also pointed out in the Introduction, the SFDM core density scales differently with size or mass, depending on the model details. More precisely, it has been known from previous analytic work and from DM-only simulations that the core mass grows with halo mass in characteristic ways that differ between FDM and SFDM-TF. As a result, the density of these cores also scales differently, where the FDM core density scales like $\rho_c \propto M_c^4 \propto M_h^{4/3}$, whereas the SFDM-TF core density scales as $\rho_c \propto M_c \propto M_h^{2/3}$. 
Consequently, if halo masses are large enough, FDM core densities may get in potential conflict with observations of dwarf galaxies, 
as pointed out in \citet{Robles_SFDM_smallscale}, who showed that FDM may do worse than CDM when it comes to typical "cusp-core galaxies". On the other hand, according to \citet{Dawoodbhoy_2021} SFDM-TF does not suffer from this conflict, and SFDM-TF fares better than CDM and FDM, \textit{if} the cores are large enough, $R_\text{TF} \gtrsim 1$ kpc. 
An obvious question to pose is how adiabatically modified SFDM-TF halos compare to those results. To this end, we plot a similar figure than Figure 11 in \citet{Dawoodbhoy_2021} showing velocity profiles, including a dataset of the same SPARC catalog by \citet{Lellietal}, but now using our models. We also include CDM for this comparison, see Figure \ref{fig:Datacomparison}.
The data we plot consists of 18 dwarf galaxies with asymptotic circular velocities between $80-100$ km s$^{-1}$, which is a range that fits with our previous choice of example "cusp-core halo" with $M_h = 10^{11}~M_{\odot}$, $R_h = 96$ kpc, and we choose these same halo parameters for our DM models. The CDM halo is parameterized with a NFW profile,
\begin{equation} \label{eq:nfw}
\rho_\text{NFW}(r) = \frac{\delta \rho_\text{crit}}{\left(\frac{cr}{R_{200}}\right)\left(1+\frac{cr}{R_{200}}\right)^2},    
\end{equation}
with the current critical background density $\rho_\text{crit} = 3H_0^2/(8\pi G)$, where we choose a Hubble parameter of $H_0 = 70$ km s$^{-1}$ Mpc$^{-1}$. (The subscripts "200" and "$h$" used to indicate global halo parameters are identical.)
Furthermore, 
\begin{equation}
    \delta = \frac{\Delta_\text{crit}}{3}\frac{c^3}{\ln (1+c) -c/(1+c)},
\end{equation}
where $\Delta_\text{crit}=200$ describes the mean overdensity of a halo compared to the background.
For our plots, we choose a concentration parameter of $c=10$, according to the median value for halos of that mass today according to the mass–concentration relation reported by \citet{Klyping2016_Massrelation}, which corresponds to a maximum circular velocity of around $v_{c,\text{max}} = 90$ km s$^{-1}$ at a radius of about $r_\text{max} = 23$ kpc. This is the same choice than \citet{Dawoodbhoy_2021} have used. 

The left panels of our Figure \ref{fig:Datacomparison} show the comparison of SFDM and CDM with data, while the right panels show the same data compared to CDM only. The models in the top panels (SFDM, CDM) have used the Blumenthal AC method, while the models in the bottom panels (SFDM, CDM) have been calculated using the Gnedin AC method. 
Let us focus first on the right panels that show adiabatically modified CDM halos with various baryon fractions. In both panels, we can see that CDM, with initial parameters appropriate for the range of velocity data shown, is not in accordance with the data. Instead, the CDM-plus-baryon halos predict much higher central velocities (and densities - not shown here). This is a post-AC exemplification of the cusp-core problem for dwarf galaxies of a total mass of $\sim 10^{11} ~M_{\odot}$.  Comparing between the results from the Blumenthal method (top right) with the bottom right panel of the Gnedin approach, we can see that the post-AC central velocities (and densities) are reduced for the latter, but it is not nearly enough to bring CDM in line with the shown data. 

Now, let us look at the left panels. Here, we plot two SFDM-TF halo models that have the same global halo parameters, but different core radius; $R_\text{TF} = 0.1$ kpc (green dashed) and $R_\text{TF} = 4$ kpc (dash-dotted curves of different color for different baryon fractions, according to the legend). Also, for comparison's sake we include the CDM model for $f=0.1$ (blue dashed) which is also shown as one case in the right panel.
We can see several things. First, the adiabatically modified SFDM-TF halo with small core exhibits a velocity profile quite close to the one of the CDM halo (compare the two dashed curves at high $v$). This is an expected result, because the smaller the SFDM core, the more similar the model to CDM, until they become indistiguishable (at least at the level of spherical density and velocity profiles). By contrast, the SFDM-TF halo with large core has much lower central velocities, regardless of the value for the baryon fraction, which clearly distinguishes it from either CDM, or from SFDM-TF with small core. Quantitative differences brought about by the various baryon fractions are mostly noticeable in the very central parts, $r \lesssim 1$ kpc, i.e. within the halo core.

However, we further notice another exemplification of the critical difference between our two employed AC routines discussed above, when it comes to large cores in SFDM. The particular SFDM-TF model with large core does quite well in reproducing the galaxy data,
as seen in the top left panel. Its core radius, $R_\text{TF} = 4$ kpc, is the same as in the model in \citet{Dawoodbhoy_2021} which is very close to ours, except that in our case we plot the AC-modified profiles using the Blumenthal method, while \citet{Dawoodbhoy_2021} plot their SFDM-only model, but modified accordingly to account for baryons \textit{post factum}. Our case also shows several curves for different baryon fraction. However, the very same SFDM-TF model, probed again with different baryon fractions, fails to reproduce the data when the Gnedin method is used, see the bottom left panel. In this case, the central velocities get reduced to a point where the model is not good any longer. Of course, we could pick new halo/core parameters to make the bottom panel of SFDM (using the Gnedin method) be in accordance with the data, but our main point here is to show that even differences in the AC routine can be decisive in judging the superiority of a cored-DM model, compared to CDM. Since our top left panel fits so well with the data, as well as with the theoretical results of \citet{Dawoodbhoy_2021}, we might jump to the conclusion that the Blumenthal AC routine is preferred. Nevertheless, we think it is necessary to study this issue in more detail, including more comparison to data. In fact, the plotted data in our figure itself exemplifies big error bars at small scales, i.e. the discrimination between models, either with respect to the DM particle parameters, or with respect to the modelling details, will be only settled with data, if they are of high quality up to the innermost regions of "cusp-core" dwarf galaxies. On the theory side, we will defer a more in-depth parameter and modelling analysis to future work.

Before we leave this subsection, we have to comment on an important shortcoming of our analysis, now that we discuss the comparison to galaxy data.
In our work, we have not included baryon feedback beyond AC. In particular, stellar feedback has been known to change the dynamics in the central parts of dwarf galaxies. Once a critical baryon density threshold is reached, star formation is inevitable. Short-lived massive stars produce stellar winds, outflows or supernovae at the end of their lives, imparting kinetic energy into the ambient gas that expands outward from the galactic centers as a result, dragging the DM along with it akin to an "inverse AC". This process is able to reduce both the DM and the baryon densities in these centers, though the details depend upon many different astrophysical processes, and the energy input into the gas has many ways to dissipate away. 
In any case, it implies that the AC-modified halo profiles of our work provide upper limit results, concerning central SFDM and baryon densities and that in reality these densities could be lower if stellar feedback is included. As such, our results can be used to estimate the amount of required stellar feedback in order to bring models in accordance with data, or conversely to find that substantial stellar feedback may not be necessary.
Looking again at the left panels of our Figure \ref{fig:Datacomparison}, we could argue that, in order to bring CDM or SFDM-TF with small core in alignment with the data, sufficient stellar feedback would be needed to reduce their central AC-modified density profiles. By contrast, SFDM-TF with large core already fits the data well (top left panel), as we pointed out previously, and there is no requirement for dynamically relevant stellar feedback. And if we compare to the bottom left panel that used the Gnedin AC approach, we could argue that stellar feedback would be certainly counterproductive for SFDM-TF with large cores, for it may reduce the central densities even further, enhancing the offset between model and data. On the other hand, if the central densities post-AC are already "low", then star formation and stellar feedback may not be efficient in the first place in order to create significant dynamical impact, anyway.
Thus, we see that a detailed study of stellar feedback in SFDM models will be required in the future to draw quantitative conclusions, concerning the ability of SFDM models to explain galaxy data satisfactory.

\subsection{Comparison to FDM}
\label{sec:54}

In this subsection, we compare our findings for AC-modified SFDM-TF profiles with the only FDM-plus-baryon halos in the literature, namely from the simulation by \cite{Mocz2020}. These FDM simulations use a "typically" small box of comoving length of $1.7 h^{-1}$ Mpc on a side and a boson mass of $2.5 \cdot 10^{-22}$ eV/$c^2$. The runs include local and global baryon physics, such as stochastic star formation, supernova feedback through kinetic winds, and primordial and metal-line cooling. 
 Furthermore, reionization happens, though uniformly and instantaneously, at a redshift of $z \sim 6$. Over time the de Broglie scale becomes increasingly computationally prohibitive to resolve.  Therefore, the cosmological simulation is stopped at $z = 5.5$.
 \cite{Mocz2020} identify some of their host halos that they find at $z = 6$, resulting from different cosmological comparison simulations that they perform: CDM, FDM (they call it BECDM), and a warm dark matter proxy simulation ("WDM") which has a cutoff in its linear power spectrum, but is otherwise just like CDM. 
 It was found that global star formation and metal enrichment in first galaxies are delayed, compared to a respective CDM-plus-baryon simulation, a result which seems not entirely attributable to the cutoff in the linear power spectrum of FDM, because the FDM runs also differ even from the proxy "WDM" run in this regard. \cite{Mocz2020} characterize the three halos that they find at the end of their simulation run, each with $M_{200} \sim 10^{10} ~M_{\odot}$.
Figure 4 in \cite{Mocz2020} shows radially averaged (comoving) density profiles for the dark matter, gas, and stars for three halos under different cosmologies at $z = 6$. The halo profiles from the DM-plus-baryon runs are quite similar to the DM-only runs (for each halo and each cosmology), which prompts \cite{Mocz2020} to conclude that "baryons have not strongly modified the dark matter potential wells for these low mass halos in the early universe". This result would suggest that the overall dynamic impact of baryons onto DM is small in high-redshift galaxies.
Furthermore, for each halo, the FDM density profiles are steeper in the centers than for CDM, or "WDM". In fact, \cite{Mocz2020} find that the \textit{smallest, densest}, most massive soliton profile is required to match the simulations of the central parts approximately; the solitonic core has a radius of $\sim 1$ kpc at $z = 6$. This last finding points right toward the issue that we discussed above, concerning the high-density cores of FDM halos, see also \citet{Robles_SFDM_smallscale} and \citet{Dawoodbhoy_2021} for details.

\begin{figure}
    \centering
    \includegraphics[width=\columnwidth]{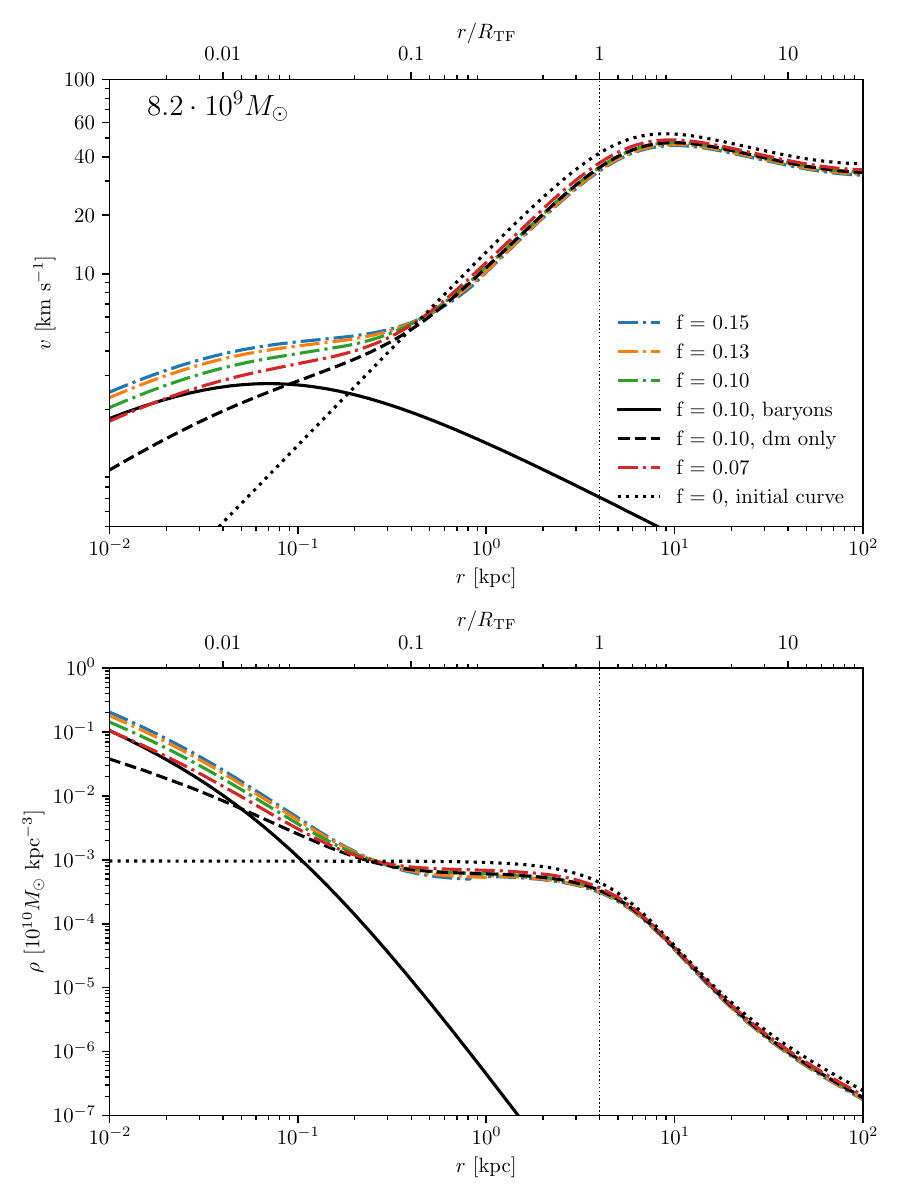}
    \caption{\textbf{Adiabatic contraction of a SFDM-TF halo} with the same global halo parameters of $M_h = 8.2 \cdot 10^9 M_\odot$ and $R_h = 42$ kpc than that of the most massive "halo 1" in the cosmological simulations of FDM performed by \citet{Mocz2020}. Our halo has a TF core radius of $R_\text{TF} = 4$ kpc. \textit{Top panel:} Circular velocity profiles. \textit{Bottom panel:} Corresponding density profiles. This plot may be compared to Figure 4 in \citet{Mocz2020}, but note our expanded $x$-axis range to smaller spatial scales. The halo envelope remains almost unchanged upon AC and looks similar to the FDM "halo 1" in \citet{Mocz2020}, while the core region experiences compression and thus a marked density increase of more than two orders of magnitude, compared to the DM-only case. We find that, in order to obtain comparable densities around $r \sim 1$ kpc between the FDM "halo 1" of \citet{Mocz2020} and our SFDM-TF halo model here, the latter requires a core radius at least four times as large as for FDM. See also caption of Fig. \ref{fig:different_core_radii}. }
    \label{fig:mocz-comparison}
\end{figure}

Of course, the high-redshift halos of \cite{Mocz2020} do not reflect the state-of-affairs of contemporary dwarf galaxies. But we may still take the FDM density profiles of \cite{Mocz2020} at face value, and check such a case in order to see how it compares to our SFDM-TF model. 
We are especially interested in the most massive FDM halo called "halo 1", see Table 1 and Figure 4 in \cite{Mocz2020}. It has a mass and radius of $M_{200} = 8.2 \cdot 10^9~M_{\odot}$ and $R_{200} = 42$ kpc, respectively. The fraction of baryonic matter in gas is $0.11$, while that in stars is $0.0057$, i.e. it has a total baryon fraction of roughly $f \lesssim 0.12$. In terms of total mass, this halo would fall in the category of "too-big-to-fail halo", rather than being a typical "cusp-core halo". Using the Blumenthal AC routine, we calculate adiabatically modified SFDM-TF halo profiles, given the above halo parameters, while we vary the TF radius so long as to come up with similar central densities than \cite{Mocz2020} find for their "halo 1". The result can be seen in our Figure \ref{fig:mocz-comparison}. While their halo contains a core with a radius of 1 kpc, we need to choose a core radius four times larger, $R_\text{TF} = 4$ kpc, in order to achieve comparable densities at a radius of roughly 1 kpc. We find that outside of this radius range, AC has little effect on the initial curve, and the envelope looks very similar to that in \cite{Mocz2020}, which confirms the argument that SFDM halo envelopes are similar to each other and to those in CDM, in the coarse-grained way discussed earlier. Again, the main changes in our SFDM-TF halo density and velocity profiles post-AC occur towards the center at radii which, unfortunately, can not be resolved by the simulations of \citet{Mocz2020}. At these smaller radii, $r \lesssim 1$ kpc, AC causes the density to rise steeply, by more than two orders of magnitude, as also observed in figures of previous subsections. In this regard, we would say that there is a dynamical impact from baryons onto SFDM in the central halo parts that were simply not resolved in \cite{Mocz2020}. There is, however, a caveat with this comparison, as follows. All of our calculations assumed a value of $\chi = 1$ for the core-envelope halo profiles. If we were to choose a smaller number, e.g. $\chi = 0.1$, we would obtain similar densities as \cite{Mocz2020} with the same core radius of $R_\text{TF} \sim 1$ kpc. Since $\chi$ is related to the characteristic radii $R_\text{TF}$ and $r_0$, see (\ref{eq:chi_parameter}), which by itself depend upon the central value of the density, a different choice of $\chi$ implies a different resulting core density. But as we stressed earlier, there are good physical reasons to fix $\chi = 1$, after all. Eventually, a quantitative comparison between FDM and SFDM-TF halo profiles will need to be based upon a more thorough investigation, also with respect to the AC approach for FDM, which we defer to future work. 

Finally, we may compare Figure \ref{fig:mocz-comparison} with the bottom panel of Figure \ref{fig:different_core_radii}, which show SFDM-TF halos of different mass, but same core radius, $R_\text{TF} = 4$ kpc, and same other parameters, as well. We can see that the central halo density shrinks, the smaller the halo mass, in accordance with relation (\ref{eq:densscale}).

\section{Conclusions}
\label{sec:conclude}

We have studied in this paper Bose-Einstein-condensed scalar field dark matter (SFDM), made of a single species of (ultra-)light bosons with mass $m\gtrsim 10^{-22}$ eV/$c^2$. In previous literature, SFDM has been found to be a promising candidate for the cosmological DM, given its potential to resolve the small-scale crisis of CDM. We focused on models with strongly repulsive self-interaction (SI) in the Thomas-Fermi regime, also called SFDM-TF. In this case, the characteristic length scale below which structure is suppressed is related to the TF radius, $R_\text{TF}$, which is much larger than the de Broglie length of the bosonic particles.
In contrast, the opposite regime of fuzzy dark matter (FDM) has as its single length scale the de Broglie length. In both cases, the genuine quantum fluid behavior of SFDM provides a means to distinguish it dynamically from CDM, if either of these length scales is large enough.

Now, the TF radius depends upon the two free parameters of the model, boson mass and 2-boson SI coupling strength, in a characteristic way. Therefore, constraints on $R_\text{TF}$ imply constraints on SFDM-TF models. We were particularly motivated by the cusp-core problem in dwarf galaxies, and in order for SFDM-TF to resolve this issue, we find that values of $R_\text{TF} \gtrsim 1$ kpc are desirable. This is in accordance with results from \citet{Dawoodbhoy_2021}. In fact, that latter work has devised a novel modelling scheme to handle the complexities of SFDM-TF halo formation, within the framework of spherical infall, and it has been established there that the halos formed that way exhibit a core-envelope structure, akin to similar findings for FDM in previous literature. In SFDM-TF, the core is close to an ($n=1$)-polytrope with radius $R_\text{TF}$, while the envelope is CDM-like. Also, it was shown there that analytic halo models with such central ($n=1$)-polytropic core supplemented with an effective isothermal sphere as halo envelope are good approximations for the accurately simulated halo profiles.
However, that work was limited to a SFDM-only analysis. In our paper here, we study new aspects by using the previously devised analytic core-envelope halo models and study adiabatic contraction (AC) of such halos due to the presence of baryons. In order to accomplish this, we had to revisit the assumptions on which the AC framework is based, foremost the notion of orbits, given that SFDM obeys different equations of motion, compared to CDM. To this end, we derived the quantum Hamilton-Jacobi equation that describes SFDM dynamics from the Hamilton dynamics perspective, which allows to find canonically conjugated action-angle variables and integrals of motion, which are the basis of the adiabatic invariants that we use. Our results apply to spherically symmetric SFDM-TF halos. We calculated example dark matter orbits and could show that the orbital structure within halo cores is little affected by the detailed choice of SFDM particle parameters.

Then, we applied our adapted AC routines to SFDM-TF halos at large, in order to study the impact of baryons onto halos of mass $\sim 10^{11}~M_\odot$, by varying the baryon fraction, as well as the core radius $R_\text{TF}$. We adopted the original method by \cite{blumenthal1986}, as well as the modification introduced by \cite{Gnedin_2004}, properly adapted to our needs. We compared our AC-modified velocity profiles to rotation curves of dwarf galaxies from the SPARC catalog in \cite{Lellietal}, and to AC-modified CDM halos parameterized with a NFW profile.
We found that SFDM-TF models with initial ("primordial") kpc-size core, $R_\text{TF} \gtrsim 1$ kpc, reproduce the data well, while SFDM-TF with sub-kpc cores, $R_\text{TF} \lesssim 0.5$ kpc, face the same issues than CDM, namely that the central velocities and densities of AC-modified halo profiles are much too high to bring them in accordance with the data. However, we found that the Blumenthal and Gnedin approaches can yield different conclusions. Using the Blumenthal method, our SFDM-TF halo models with large core not only fit with data, but also with the models in \citet{Dawoodbhoy_2021}. On the other hand, it has been known that the Gnedin approach leads to a reduction in the density enhancement post-AC of roughly a factor of two, compared to the Blumenthal approach.
We also confirm this for our models of SFDM-TF with small cores and CDM.
However, we observe a significant reduction in the central values for velocity and density for SFDM-TF halos with large primordial core, to an extent that the same model cannot reproduce the data anymore. A more detailed parameter study and more comparison to data will be necessary in the future to clarify how robust such differences - brought about by different AC routines - really are. In any case, it points to the fact that more work is required in order to disentangle the various impacts caused by SFDM parameters versus those caused by the dynamical modeling approach.
We also compared AC-modified halos to some findings of the cosmological FDM-plus-baryon simulations by \cite{Mocz2020}, which produced few halos of mass $\sim 10^{10}~M_{\odot}$ within their comoving simulation box of roughly 2 Mpc on a side. We found that, in order to have the same halo densities at a radius of roughly 1 kpc than those formed in these simulations, we require core radii roughly 4 times larger than those FDM halo cores found in \cite{Mocz2020}. However, these simulations cannot resolve the innermost parts of halos the same way than our models do, i.e. a cleaner comparison to our results will have to wait for an extension of our AC models to the FDM regime, which is beyond the scope of this paper. 

A shortcoming of our work concerns the fact that we have not included baryon feedback beyond AC. Once a critical threshold in baryon density is reached, star formation is inevitable, and the ensuing stellar feedback of massive stars is able to impart kinetic energy into the ambient gas. As a result, it will expand outward from the galactic centers, dragging the DM along with it. This process is able to reduce both the DM and the baryon densities in these centers, though the efficiency depends upon many astrophysical processes and parameters. Therefore, the AC-modified halo profiles of our work provide upper limit results, concerning central SFDM and baryon densities and velocities, and in reality these could be lower if stellar feedback is included. At face value, our results can be used to estimate the amount of required stellar feedback in order to bring models in accordance with data, or conversely to find that substantial stellar feedback may not be necessary. Currently, our results suggest that typical "cusp-core" SFDM-TF halos of mass $\sim 10^{11}~M_{\odot}$ having large cores, $R_\text{TF} \gtrsim 1$ kpc, may not require stellar feedback, and "purely" AC-modified profiles are sufficient to reproduce observational data of dwarf galaxies. However, more future work will be required to draw final conclusions.
 
We stress that the modeling of the various impacts of baryons within both CDM and non-CDM models remains a hot topic, which is more advanced for some models, compared to others. One example is SIDM\footnote{Roughly speaking, SIDM is a model where a finite cross section, which can also depend upon the velocity, is added to "CDM". As such, SIDM exhibits differences to CDM on (dwarf-)galactic scales, while being CDM-like on larger and cosmological scales. In particular, SIDM predicts central galactic cores, like SFDM.}. Detailed hydrodynamical simulations that include a host of baryon physics have not yet converged to a definite answer. For instance, \cite{Robles_etal_SIDM} find that central SIDM profiles of halos with mass $\sim 10^{10}~M_{\odot}$ remain flatter compared to CDM counterparts. Studying the same halo masses, \cite{Fry_etal_SIDM} report that SIDM-plus-baryon profiles are effectively indistinguishable from CDM-plus-baryon profiles. Recently, \cite{Sameie_etal_SIDM} have found that SIDM halos of Milky Way size $\sim 10^{12}~M_{\odot}$ can reach \textit{higher} central densities than corresponding CDM counterparts, as a result of baryon feedback.  
  Detailed simulations for SFDM will be required to address the same questions in a quantitative manner, as emphasized already.    
   Taken together and in light of our findings, we might conclude that, in general, non-CDM models with parameters that produce too small, "compact" primordial cores are subject to the same issues than CDM, when it comes to the cusp-core problem.

In addition, we call for the necessity to obtain more high-quality observational rotation curves of dwarf galaxies with halo mass of $\sim 10^{10}-10^{11}~M_{\odot}$, especially down to very small spatial scales, in order to be able to disentangle features in halo profiles that are caused by DM microphysics from those due to the dynamical modelling approach. As a matter of fact, many of the decisive features between models just occur in the central halo parts, or close to the transition between halo core and envelope. Therefore, the comparison to more data is highly warranted. 
On the theory side, we will defer to future work a detailed parameter study of our current AC models, as well as an extension of them to cover the FDM regime. Furthermore, future investigations require the addition of stellar feedback in order to determine quantitatively how much of such feedback on top of the AC-modified halo profiles is actually necessary to resolve the cusp-core problem of dwarf galaxies within the SFDM paradigm.

Finally, we comment on an important implication that results from the recent work by \citet{Shapiro2021}, whose conclusions were confirmed subsequently by \cite*{Hartman_2022_constraints} and \cite{Foidl}. By extending the SFDM-TF halo formation studies of \citet{Dawoodbhoy_2021} and by performing a semi-analytic linear structure formation calculation, it has been found in \citet{Shapiro2021} that the (unconditional) halo mass function in SFDM-TF exhibits a cutoff at higher halo mass than the corresponding one for FDM, although the subsequent falloff toward smaller masses is much shallower than in FDM.  The upshot is that SFDM-TF halos with kpc-size primordial cores are highly constrained, and it strongly appears that sub-kpc primordial cores, $R_\text{TF} \lesssim 0.1$ kpc, are favored, instead. This implies that we would need to limit our model parameters to those that produce sub-kpc cores, and we have seen in this work here that these models produce AC-modified halo profiles close to CDM. The combined findings of the above-mentioned previous works along with our studies here would then suggest that SFDM-TF may require stellar feedback to resolve the cusp-core problem, after all. Again, more future work will be required to settle this question definitely.

\section*{Acknowledgements}
We thank Ryan Leaman and Glenn van de Ven for helpful discussions. The authors acknowledge the support by the Austrian Science Fund FWF through an Elise Richter fellowship, grant nr. V 656-N28, to T. Rindler-Daller. 


\section*{Data Availability Statement}
All data are incorporated into the article.



\bibliographystyle{mnras}
\bibliography{references}




\appendix

\section{Properties of the TF halo core as an ($\lowercase{n}=1$)-polytrope}
\label{appendix:thomas-fermi}

Here we outline the calculation of the density profile for the TF core, equation \eqref{eq:polytrope-density}, starting from the differential equation \eqref{eq:TF-regime-diffeq}. From this profile we can subsequently derive the analytic expressions of the gravitational potential, the mass profile and the velocity profile.

\begin{figure*}
    \centering
    \includegraphics[width=0.33\textwidth]{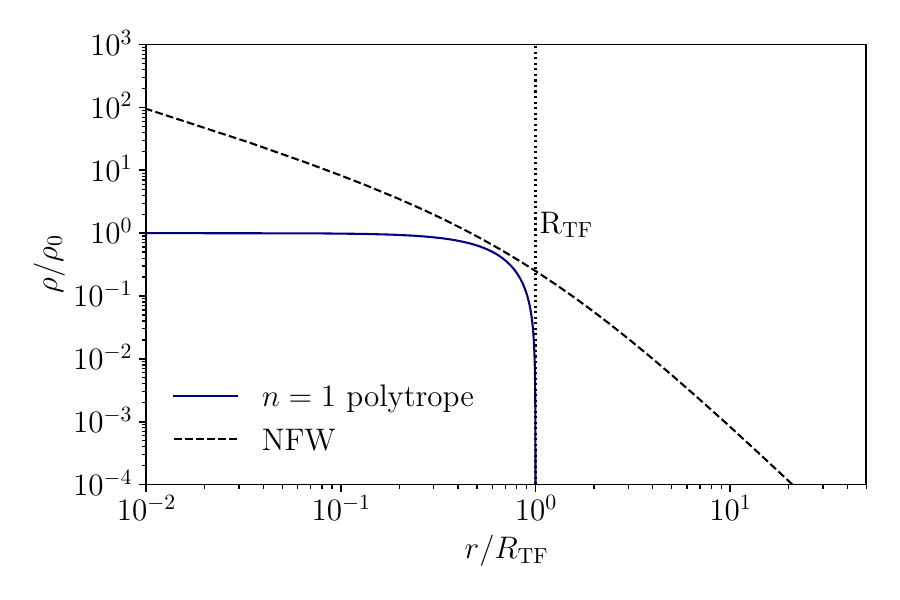}
    \includegraphics[width=0.33\textwidth]{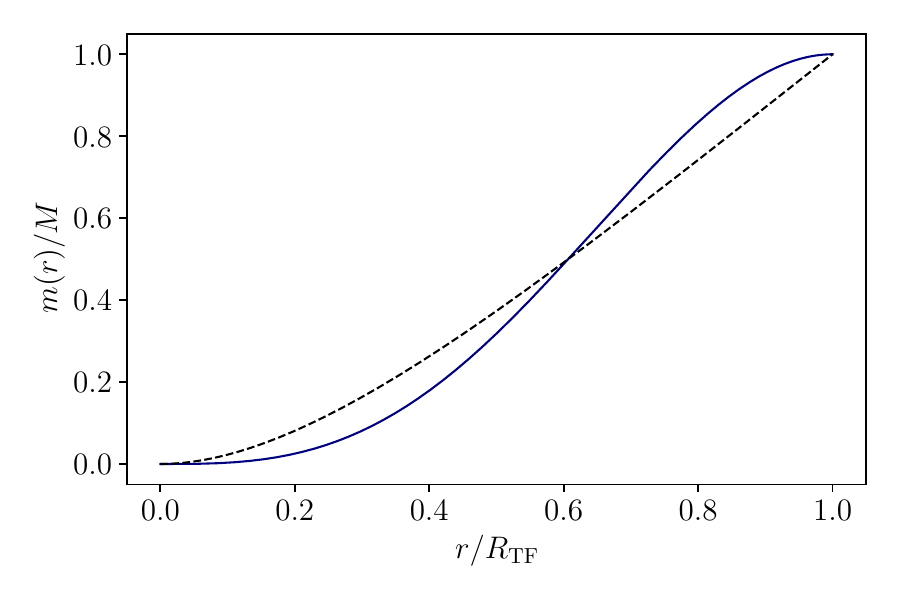}
    \includegraphics[width=0.33\textwidth]{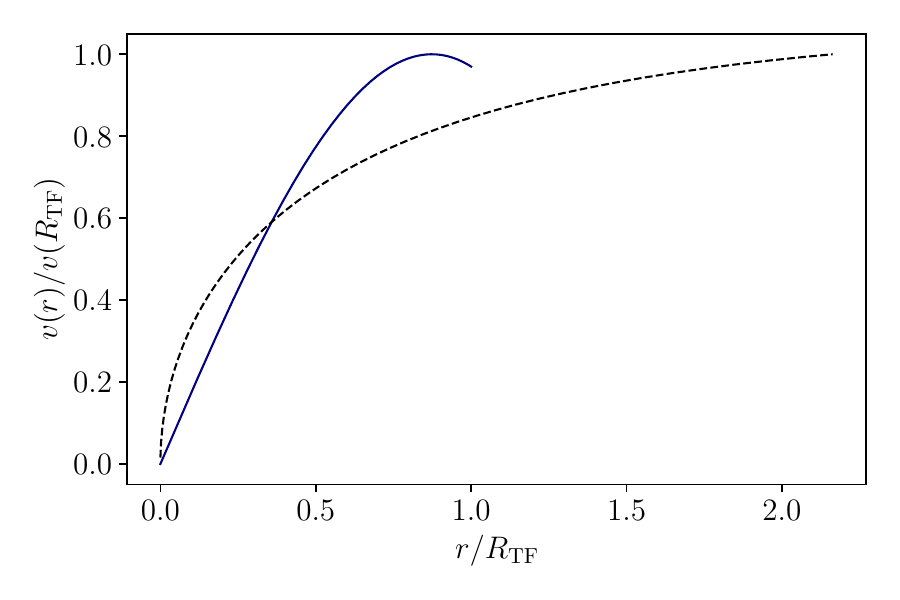}
    \caption{\textit{Left panel}: Normalized density profile for an ($n=1$)-polytrope, representing a TF halo core, and an NFW profile, for illustrative purposes plotted together. Here, the TF radius $R_\text{TF}$ acts as the scale radius for the NFW profile. Additionally, we use a concentration parameter of $c = 1$. \textit{Center panel}: Normalized enclosed mass profiles for the same configuration. \textit{Right panel}: Normalized circular velocity for the same configuration. The polytrope reaches its maximum velocity at $r \approx 0.85 R_\text{TF}$, while the NFW profile has its maximum farther outside at $r \approx 2.163 R_\text{TF}$.}
    \label{fig:polytrope}
\end{figure*}

We consider spherical symmetry. After taking the derivative and multiplying by $r^2$, equation \eqref{eq:TF-regime-diffeq} reads
\begin{align*}
    -\frac{1}{r^2} \frac{\dd}{\dd r} \left( r^2 \frac{g}{m^2} \frac{\dd \rho}{\dd r} \right) = 4 \pi G \rho,
\end{align*}
where Poisson's equation (\ref{eq:poisson2}) and the definition of the SI-pressure (\ref{eq:polytropic-pressure}) were used. With help of the substitutions
\begin{align}
    r = \alpha z \quad \text{ and } \quad \dd r = \alpha \dd z,
\end{align}
for the radius, where
\begin{align*}
    \alpha = \left( \frac{n + 1}{4 \pi G} K_\rho \rho_0^{1/n - 1} \right)^{1/2} = \left( \frac{K_\rho}{2 \pi G} \right)^{1/2},
\end{align*}
and
\begin{align}
    \rho (r) = \rho_0 \theta^n (z) = \rho_0 \theta(z)
\end{align}
for the density, setting $n=1$, the above differential equation can be further simplified, so that we arrive at
\begin{align}
    \frac{1}{z^2} \frac{\dd}{\dd z} \left( z^2 \frac{\dd \theta}{\dd z} \right) + \theta = 0.
\end{align}
This is the familiar Lane-Emden equation for self-gravitating polytropic fluids with polytropic index $n = 1$. It is accompanied with the boundary conditions $\theta(0) = 1$ and $\theta'(0) = 0$ to guarantee finite central halo densities. With another substitution,
\begin{align}
    \xi(z) = z \theta, \quad \frac{\dd z}{\dd \theta} = \frac{\xi' z - \xi}{z^2},
\end{align}
where the prime denotes the derivative with respect to $z$, we arrive at the differential equation
\begin{align}
    \xi'' + \xi = 0,
\end{align}
for which the general solution is given by
\begin{align}
    \theta(z) = a \frac{\sin(z)}{z} + b \frac{\cos(z)}{z}.
\end{align}
The constants $a, b$ are then easily determined with the boundary conditions, and after backward substitution the density profile results,
\begin{align} \label{eq:appendix-polytrope-density}
    \rho(r) = \rho_0 \frac{\sin (\pi r / R_\text{TF})}{\pi r / R_\text{TF}},
\end{align}
where we also made use of the polytropic constant $K_\rho$ as defined in equation \eqref{eq:polytropic-pressure} in order to express the density in terms of the TF radius (\ref{eq:tfradius}). For purely illustrative purposes, we provide a comparison of this profile to an NFW profile with $c = 1$ and $R_\text{TF}$ as scale radius in Figure \ref{fig:polytrope}.

With the expression for the density in (\ref{eq:appendix-polytrope-density}), we can now use Poisson's equation to derive the gravitational potential. We obtain 
\begin{align}
    \frac{\dd \Phi}{\dd r} &= \frac{4 G \rho_0 R_\text{TF}}{r^2} \left( \frac{\sin(\pi r / R_\text{TF}) - \pi r / R_\text{TF} \cos (\pi r / R_\text{TF})}{\pi^2 / R_\text{TF}^2} + c_1, \right)
\end{align}
hence,
\begin{align}
    \Phi(r) &= \frac{4 G \rho_0 R_\text{TF}^3}{\pi^2} \left( -\frac{\sin(\pi r / R_\text{TF})}{r} - \frac{c_1 \pi^2 / R_\text{TF}^2}{r} + c_2 \right),
\end{align}
for which we again need to determine the integration constants. We employ the usual boundary conditions, namely no net force at the center, $\dd \Phi/ \dd r = 0$ at $r=0$, and a point mass at infinity, $\Phi(r) = - 1/r$ for $r \to \infty$, which finally leads to the gravitational potential of an ($n=1$)-polytrope,
\begin{align}
    \Phi(r) = - \frac{4 G \rho_0 R_\text{TF}^3}{\pi^2} \frac{\sin(\pi r / R_\text{TF})}{r}.
\end{align}
From the density profile \eqref{eq:appendix-polytrope-density}, we can further derive the enclosed mass $M(r)$ inside a given radius $r$, which is
\begin{align}
    M(r) &= \int_0^r 4 \pi r'^2 \rho(r') \dd r' \\
    &= \frac{4 \rho_0 R_\text{TF}^3}{\pi^2} \left( \sin(\pi r / R_\text{TF}) - \frac{\pi r}{R_\text{TF}} \cos(\pi r / R_\text{TF}) \right).
\end{align}
Finally, we calculate the circular velocity profile as
\begin{align}
    v(r) &= \sqrt{\frac{GM(r)}{r}} \\
    &= \sqrt{\frac{4 G \rho_0 R_\text{TF}^2}{\pi} \left( \frac{R_\text{TF}}{\pi r} \sin(\pi r / R_\text{TF}) - \cos(\pi r / R_\text{TF}) \right)}.
\end{align}
Figure \ref{fig:polytrope} provides comparisons to a NFW profile for the enclosed mass and circular velocity profiles, respectively, for illustrative purposes.

\section{Non-dimensionalizing the core-envelope equation: "method 2"} \label{appendix:core_envelope_method_2}

\citet{Chavanis_2019} and \citet{Dawoodbhoy_2021} use the (standard) exponential ansatz for the density, as well as the characteristic radius $r_0$ of the isothermal sphere to non-dimensionalize length scales. Here, in "Method 2" we use the substitutions
\begin{align*}
    \Tilde{\rho} = \frac{\rho}{\rho_0} \quad \quad \text{and} \quad \quad \zeta = \frac{r}{R_\text{TF}}
\end{align*}
in order to non-dimensionalize equation \eqref{eq:core_envelope_1}; in particular we use $R_\text{TF}$, instead of $r_0$. We further use the fact that the circular velocity at radius $R_\text{TF}$ is given by
\begin{align*}
    v^2 (R_\text{TF}) = \frac{GM_c}{R_\text{TF}},
\end{align*}
where $M_c$ denotes the enclosed mass at this radius, i.e.
\begin{align*}
    M_c \equiv M (R_\text{TF}) = \frac{4 \rho_0 R_\text{TF}^3}{\pi},
\end{align*}
which is the mass of the TF core.
Again, a parameter is introduced that describes the dominance of the TF core compared to the envelope, in this case 
\begin{align}
    \kappa = \frac{\sigma^2}{v_c^2} = \frac{1}{\chi}.
\end{align}
Equation \eqref{eq:core_envelope_1} can thus be written as
\begin{align} \label{eq:core_envelope_3}
    \frac{\dd}{\dd \zeta} \left( \kappa \zeta^2 \frac{\dd \ln \tilde{\rho}}{\dd \zeta} + \zeta^2 \frac{\dd \tilde{\rho}}{\dd \zeta} \right) = -\pi^2 \zeta^2 \tilde{\rho}
\end{align}
and solved numerically with the same boundary conditions of finite central density, $\tilde{\rho}(0) = 1$ and $\tilde{\rho}'(0) = 0$, as used previously. 
A comparison of the solutions of this differential equation, called "Method 2", and the solutions to equation \eqref{eq:core_envelope_2}, upon the approach by \citet{Chavanis_2019}, is provided in Figure \ref{fig:core-envelope-comparison} in the main text.

\section{St\"ackel conditions for spherical SFDM/BEC-DM halos} 
\label{appendix:staeckel-conditions}

In this appendix, we study the St\"ackel conditions \cite{staeckel}, as listed e.g. in \citet{bookGoldstein}, and check them for the Hamiltonian in equation \eqref{eq:hamiltonian},
\begin{align} \label{eq:hamiltonian-appendix}
    \mathcal{H} &= \frac{\mathbfit{p}^2}{2m} + mQ + m\Phi + \frac{g\rho}{m} \nonumber \\
    &= \frac{1}{2m} \left( p_r^2 + \frac{p_\vartheta^2}{r^2} + \frac{p_\varphi^2}{r^2 \sin^2\vartheta} \right) + mQ + m\Phi + \frac{g\rho}{m}.
\end{align}
The five conditions are
\begin{enumerate}
    \item The Hamiltonian is conserved.
    \item Moreover, it can be written as
    \begin{align} \label{eq:second-staeckel}
        H = \frac{1}{2} \left( \mathbfit{p}^\top - \mathbfit{a}^\top \right) \mathcal{T}^{-1} (\mathbfit{p} - \mathbfit{a}) + V(q),
    \end{align}
    i.e. the Lagrangian of the system is at most a quadratic function of the generalized velocities. $\mathcal{T}$ is a square matrix whose elements depend on the chosen coordinate system.
    
    \item The elements of $\mathbfit{a}$ depend only on the corresponding generalized coordinate, i.e. $a_i = a_i (q_i)$.
    
    \item The potential can be written in the form
    \begin{align} \label{eq:fourth-staeckel}
        V(q) = \sum_i \frac{V_i (q_i)}{\mathcal{T}_{ii}}.
    \end{align}
    
    \item There exists a matrix $\phi$ with elements
    \begin{align} \label{eq:fifth-staeckel}
        \phi_{ij} = \frac{\partial W_i}{\partial q_i} \frac{\partial^2 W_i}{\partial q_i \partial \alpha_j}
    \end{align}
    and an inverse $\phi^{-1}_{ij}$, where the diagonal elements of both matrices are either constants, or depend only on the corresponding coordinate, i.e. $\phi_{ii}$ and $\phi^{-1}_{ii}$ depend only on the coordinate $q_i$ \citep{greenwood}.
\end{enumerate}

The first condition is satisfied, because the Hamiltonian does not explicitly depend on time. We resume with the second condition, for which equation \eqref{eq:hamiltonian-appendix} is written in matrix notation, 
\begin{align*}
    \mathcal{H} = \begin{pmatrix} p_r & p_\vartheta & p_\varphi \end{pmatrix} \begin{pmatrix} 1/m & 0 & 0 \\ 0 & 1/mr^2 & 0 \\ 0 & 0 & 1/mr^2\sin^2\vartheta \end{pmatrix} \begin{pmatrix} p_r \\ p_\vartheta \\ p_\varphi \end{pmatrix} + V(q),
\end{align*}
where $V(q)$ encapsulates the sum of quantum and gravitational potential and the SI term,
\begin{align*}
    V(q) = mQ(q) + m\Phi(q) + \frac{g\rho(q)}{m}.
\end{align*}
Here we identify 
\begin{align*}
    \mathcal{T}^{-1} = \begin{pmatrix} 1/m & 0 & 0 \\ 0 & 1/mr^2 & 0 \\ 0 & 0 & 1/mr^2\sin^2\vartheta \end{pmatrix},
\end{align*}
and a comparison with equation \eqref{eq:second-staeckel} reveals that $\mathbfit{a} = \mathbfit{0}$ for this system. Hence, conditions (ii) and (iii) are satisfied. From the constraints on the potential in equation \eqref{eq:fourth-staeckel} we require 
\begin{align*}
    V(q) = V_r(r) + \frac{V_\vartheta(\vartheta)}{r^2} + \frac{V_\varphi(\varphi)}{r^2 \sin^2 \vartheta}.
\end{align*}
Given that we assume spherically-symmetric potential-density pairs throughout this work, we have $V_\vartheta (\vartheta) = V_\varphi (\varphi) = 0$, so the fourth St\"ackel condition is automatically fulfilled. We are left with the task of finding the matrix $\phi$ for condition (iv). Since the Hamiltonian is not explicitly time-dependent, its principle function can be written in the form
\begin{align} \label{eq:principle-function}
    S(\mathbfit{q}, t) = W(\mathbfit{q}) - \alpha_t t,
\end{align}
where $W$ now denotes Hamilton's characteristic function and $\alpha_t$ denotes some (integration) constant which we immediately identify as the total energy of the system, $\alpha_t = E$. In order to calculate \eqref{eq:fifth-staeckel} we further separate this function such that
\begin{align*}
    W = W_j (q_j, \alpha) + W' (q_i, \alpha),
\end{align*}
where $W'$ depends on all generalized coordinates $q_i$, except for $q_j$. The Hamilton-Jacobi equation can then be formulated as
\begin{align*}
    \mathcal{H} \left( q_i, \frac{\partial W'}{\partial q_i}, f \left( q_j, \frac{\partial W_j}{q_j} \right) \right) = \alpha,
\end{align*}
and further inverted to result in
\begin{align*}
    f \left( q_j, \frac{\partial W_j}{\partial q_j} \right) = g \left( q_i, \frac{\partial W'}{\partial q_i}, \alpha \right).
\end{align*}
Note that $\alpha$ again represents some constant here. The left-hand-side of this equation depends only upon the coordinate $q_j$, while the right-hand-side depends on all other $q_i$, so equality can only be true, if both sides equal the same constant,
\begin{align*}
        f \left( q_j, \frac{\partial W_j}{\partial q_j} \right) = \alpha_j =  g \left( q_i, \frac{\partial W'}{\partial q_i} \right).
\end{align*}
We make use of this fact to separate Hamilton's principle function \eqref{eq:principle-function} to obtain
\begin{align*}
    S(\mathbfit{q}, t) = W_r(r) + W_\vartheta (\vartheta) + W_\varphi (\varphi) - Et,
\end{align*}
which leaves us with the Hamilton-Jacobi equation
\begin{align*}
    \frac{1}{2m} \left( \frac{\partial W_r}{\partial r} \right)^2 + mQ + m\Phi + \frac{g\rho}{m} &+ \frac{1}{2mr^2} \left(\frac{\partial W_\vartheta}{\partial \vartheta} \right)^2 \\
    &\quad \quad+ \frac{1}{2mr^2 \sin^2 \vartheta} \left( \frac{\partial W_\varphi}{\partial \varphi} \right)^2 = E.
\end{align*}
Following the procedure outlined above for the cyclic coordinate $\varphi$, another constant
\begin{align*}
    p_\varphi = \frac{\partial W_\varphi}{\partial \varphi} = \alpha_\varphi,
\end{align*}
is recognized. The remaining equation,
\begin{align*}
    \frac{1}{2m} \left( \frac{\partial W_r}{\partial r}\right)^2 + mQ + m\Phi + \frac{g\rho}{m} + \frac{1}{2mr^2} \left[ \left( \frac{\partial W_\vartheta}{\partial \vartheta}\right)^2 + \frac{\alpha_\varphi^2}{\sin^2 \vartheta} \right] = E,
\end{align*}
can once again be simplified, leading to the last constant,
\begin{align*}
    \left( \frac{\partial W_\vartheta}{\partial \vartheta} \right)^2 + \frac{\alpha_\varphi^2}{\sin^2 \vartheta} = \alpha_\vartheta^2,
\end{align*}
or in terms of the (constant) momentum,
\begin{align*}
    p_\vartheta = \frac{\partial W_\vartheta}{\partial \vartheta} =   \sqrt{\alpha_\vartheta^2 - \frac{\alpha_\varphi^2}{\sin^2 \vartheta}},
\end{align*}
which is the final piece needed to set up the sought-after matrices. Finally, applying equation \eqref{eq:fifth-staeckel} to Hamilton's characteristic function gives
\begin{align*}
    \phi = \begin{pmatrix} m & -\alpha_\vartheta/r^2 & 0 \\ 0 & \alpha_\vartheta & -\alpha_\varphi/\sin^2 \vartheta \\ 0 & 0 & \alpha_\varphi \end{pmatrix}
\end{align*}
and
\begin{align*}
    \phi^{-1} = \begin{pmatrix} 1/m & 1/mr^2 & 1/mr^2\sin^2\vartheta \\ 0 & 1/\alpha_\vartheta & 1/\alpha_\varphi \sin^2\vartheta \\ 0 & 0 & 1/\alpha_\varphi \end{pmatrix},
\end{align*}
which satisfy the fifth condition. The Hamilton-Jacobi equation \eqref{eq:qhje} is thus separable.


\bsp	
\label{lastpage}
\end{document}